\documentclass[11pt,a4paper]{article}

\usepackage{jcappub}

\usepackage{gensymb}

\newcommand{\md}{\mathrm{d}}

\title{Dynamical Dark Energy in Minimally Modified Gravity}
\author{Alexander Ganz}
\affiliation[a]{ Faculty of Physics, Astronomy and Applied Computer Science, Jagiellonian University, 30-348 Krakow, Poland}

\emailAdd{alexander.ganz@uj.edu.pl}

\abstract{ Minimally modified gravity is a class of models with only the two tensor degrees of freedom as in general relativity. Using the framework with auxiliary constraints these models can maintain a dynamical cosmological background. The form of the constraints is thereby restricted by the requirement of dynamical dark energy and the avoidance of a breakdown of perturbation theory. Studying the linear perturbations around the FLRW background the results are, however, quite insensitive to the details of the constraints leading to a modified effective gravitational constant or a non-vanishing sound speed for dust. 

}

\arxivnumber{}

\notoc 

\begin{document}

\maketitle
\flushbottom

\pagebreak

\section{Introduction}

In the recent years minimally modified gravity models (MMG) have received an increased attention \cite{Aoki:2018zcv,Lin:2018mip,Iyonaga:2018vnu,Gao:2019twq,Lin:2017oow,Mukohyama:2019unx,DeFelice:2020eju,Yao:2020tur,Tasinato:2020fni,Aoki:2021zuy,Lin:2020nro,Ganz:2021hmp,Bartolo:2021wpt}. These are a class of modified gravity models which do not add additional degrees of freedom as it is the normally the case as in scalar-tensor theories (see \cite{Frieman:2008sn,Clifton:2011jh,Joyce:2014kja} for a review).

The first model in this context has been the cuscuton model \cite{Afshordi:2006ad} where due to the peculiar kinetic term the scalar field becomes non-dynamical as long as $\partial_i \varphi=0$ \cite{Gomes:2017tzd}. However, for a non homogeneous profile of the scalar field there will be additional instantaneous modes which are, in general, unstable. These modes are not unique for MMG models but also occur in higher order scalar-tensor theories \cite{Crisostomi:2017ugk,DeFelice:2018ewo,Ganz:2019vre} or Horava-Lifshitz gravity models \cite{Blas:2009yd}. However, in \cite{DeFelice:2018ewo} it has been argued that these modes are superficial and can be removed by imposing proper boundary conditions. 

On the other hand, we could fix the slicing of the manifold from the start by choosing $\varphi=t$. In this case the full diffeomorphism invariance is explicitly broken evading, therefore, the Lovelock theorem \cite{doi:10.1063/1.1665613,doi:10.1063/1.1666069}. These models are commonly called spatial covariant gravity models (SCG). While SCG theories, in general, have three degrees of freedom (2 tensor and 1 scalar mode) due to the breaking of diffemorphism invariance \cite{Gao:2014soa,Gao:2014fra} by requiring additional degeneracy conditions on the Lagrangian as it is the case for Cuscuton the additional scalar degree of freedom can be removed (see for instance \cite{Gao:2019twq,Lin:2020nro} for a detailed discussion). Another way has been proposed in \cite{Yao:2020tur}. Instead of requiring degeneracy conditions on the form of the Lagrangian one instead imposes additional constraints by hand at the Hamiltonian level. The Lagrangian can then afterwards be obtained by performing a Legendre transformation. In our paper we will focus on this specific approach. 

A common motivation of modified gravity is to explain dark energy which is responsible for the accelerated expansion of the Universe \cite{SupernovaSearchTeam:1998fmf,SupernovaCosmologyProject:1998vns}. Modified gravity theories like scalar-tensor theories provide a dynamical degree of freedom at the FLRW background even in the absence of additional matter and, therefore, allow for a dynamical evolution of dark energy. This is not the case in general relativity (GR) where due to the high symmetries of the FLRW metric in the absence of matter the Hubble parameter is completely fixed by the cosmological constant and there is no dynamics. This is also the case for conventional MMG models where the Hubble parameter is fixed by generically time dependent functions but without a dynamical degree of freedom. Depending on the specific model this can be sufficient to model any evolution of the Hubble parameter as in VCDM \cite{DeFelice:2022uxv} by tuning the free time functions appropriately. This is, however, not generically the case since the free functions can be constrained by consistency relations \cite{Afshordi:2007yx,DeFelice:2022uxv}. Recently, it has been realized that it is also possible to construct MMG which have a dynamical degree of freedom at the background even in the absence of an additional matter content  by imposing auxiliary constraints \cite{DeFelice:2015hla,DeFelice:2015moy,Ganz:2021hmp} which vanish trivially at the background level. Therefore, it is possible to obtain the same standard background evolution as in common scalar tensor theories. Both these approaches have the advantage that is possible to model background evolution which are commonly plagued by instabilities \cite{Vikman:2004dc} as bounces \cite{Boruah:2018pvq,Kim:2020iwq} or phantom dark energy \cite{Ganz:2021hmp}. 

In this paper the focus is on the construction of MMG with a dynamical background. Imposing auxiliary constraints by hand has got an increasing attention in the literature as it provides an easy and straightforward way to obtain a MMG model. Therefore, it is important to understand possible impacts of the constraints in more detail. As we will see these models are relatively insensitive to the details of the constraints at the linear level allowing for a systematic discussion of the phenomenological properties.

As another aspect, by using the method of auxiliary constraints it is also possible to construct models which have more degrees of freedom at the linear level than at the full non-linear level signalizing a breakdown of perturbation theory. This clarifies a common misconception that linear perturbation theory allows to provide a lower limit on the total number of degrees of freedom at the full non-linear level avoiding a complete Hamiltonian analysis. In the appendix there is also a short discussion how it is possible to obtain the same features without the need of auxiliary constraints. But these models are, in general, quite pathological. 

The structure of the paper is as follows. In the first section \ref{sec:Toy_model}, we present the main idea for a toy model and discussing the fundamental properties in the Hamiltonian and Lagrangian formulation. Further, we analyze the properties of gravitational waves around a generic background. In section \ref{sec:Hamiltonian_construction} we review the approach developed in \cite{Yao:2020tur} briefly and discuss the conditions on the form of the constraints and a possible breakdown of perturbation theory. 
Using the developed framework we discuss in \ref{sec:Effective_Field} the phenomenological consequences by studying the linear perturbations around FLRW for a broad class of models. In particular, we show that for specific classes of constraints the results are relatively insensitive to the details of the constraints allowing for a systematical exploration.
Last, we shortly discuss our results and provide an outline (sec. \ref{sec:Conclusion}). 

In the paper we are using the mostly plus $(- + + +)$ signature and use units where the reduced Planck mass $M_p^2=(8\pi G)^{-1}$ is set to one.

\section{Toy model}
\label{sec:Toy_model}

In order to present the main idea let us first consider a toy model. By using the ADM decomposition
\begin{align}
    \md s^2 = - N^2 \md t^2 + h_{ij} \left( \md x^i + N^i \md t \right) \left( \md x^j + N^j \md t \right)
\end{align}
where $N$, $N^k$ and $h_{ij}$ are the lapse, the shift vector and the 3-dimensional metric on the spatial hypersurface of constant time,
the Lagrangian of K-essence can be written in the unitary gauge $\varphi=\varphi(t)$ as
\begin{align}
\label{eq:Toy_model}
    \mathcal{L} = \frac{1}{2} \sqrt{h} N  \left( K_{ij} K^{ij} - K^2 + R \right) + \sqrt{h} N P(X) ,
\end{align}
where  $R$ is the three dimensional Ricci scalar, $X=\dot \varphi^2/(N^2)$ and  $K_{ij}$ the extrinsic curvature
\begin{align}
    K_{ij} = \frac{1}{2N} \left( \dot h_{ij} - D_i N_j - D_j N_i\right).
\end{align}
with $D_j$ being the covariant dervative with respect to the spatial metric $h_{ij}$. Note, that $\dot \varphi^2(t)$ is just a function of time and not a free variable. We could further fix it by using the remaining time-reparametrization invariance by setting $\dot \varphi=1$.
The corresponding Hamiltonian is given by
\begin{align}
    H_T = \int \md^3x\, \mathcal{H} + N^k \mathcal{H}_k + u_N \pi_N + u_i \pi^i
\end{align}
where
\begin{align}
    \mathcal{H} =& \frac{2 N }{\sqrt{h}} \left( \pi_{ij} \pi^{ij} - \frac{1}{2} \pi^2 \right) - \frac{1}{2} \sqrt{h} N R - \sqrt{h} N P, \\
    \mathcal{H}_k =& -2 D_j \pi^j_k, 
\end{align}
and $\pi_N$, $\pi^i$ and $\pi^{ij}$ are the canonical conjugate momenta to $N$, $N_i$ and $h_{ij}$.

Due to the spatial diffeomorphism invariance $\mathcal{H}_k$ and $\pi^k$ are the usual 6 first class constraints corresponding to spatial transformation. On the other hand, since we fixed $\varphi=\varphi(t)$, the Hamiltonian constraint
\begin{align}
    \mathcal{H}_0 = - \{ \pi_N, H_T \} = \mathcal{H}_{GR} - \sqrt{h} P + \sqrt{h} \frac{2 \dot \varphi^2}{N^2} P_X
\end{align}
is not anymore a first class constraint but instead forms with $\pi_N$ a pair of second class constraint resulting in three degrees of freedom.

Our aim is to eliminate the scalar degree of freedom. Following the discussion in \cite{Yao:2020tur} the easiest way is to add a constraint at the Hamiltonian
\begin{align}
    H_T = \int \md^3x\, \mathcal{H} + N^k \mathcal{H}_k + u_N \pi_N + u^k \pi_k + \lambda C.
\end{align}
For simplicity let us choose 
\begin{align}
    C = \sqrt{h} D_k D^k R.
\end{align}
The consistency relation for the new constraint enforces a secondary constraint
\begin{align}
    C_R^{(2)} \approx & - 4 R^{ij} \pi_{ij} D^2 N + 2 R \pi D^2 N - \pi D_k R D^k N - 8 \sqrt{h} D_k \left( \frac{\pi_{ij}}{\sqrt{h}} \right) R^{ij} D^k N  \nonumber \\
    & + \sqrt{h} D_k \left( \frac{\pi}{\sqrt{h}} \right) \left( 4 R D^k N - N D^k R \right) + 4 \sqrt{h} N D_i D_j D^2 \left( \frac{\pi^{ij}}{\sqrt{h}}\right) + 2 \sqrt{h} N R D^2 \left( \frac{\pi}{\sqrt{h}} \right) \nonumber \\
    & + 8 \sqrt{h} D^k N  D_j D^2\left( \frac{\pi_k^j}{\sqrt{h}}\right) + 4 \pi^{ij} D_i N D_j R + 16 \sqrt{h} D_i D_j N D_k D^j \left( \frac{\pi^{ik}}{\sqrt{h}} \right) \nonumber \\
    & + 8 \sqrt{h} D^k N D_i D_j D_k \left( \frac{\pi^{ij}}{\sqrt{h}} \right) + 4 \sqrt{h} D^2 \left( \frac{\pi^{ij}}{\sqrt{h}} \right) \left( D_i D_j N - N R^{ij} \right) + 4 \pi^{ij} D_i D_j D^2 N \nonumber \\
    & + 8 \sqrt{h} D_k \left( \frac{\pi^{ij}}{\sqrt{h}} \right) D_i D_j D^k N
\end{align}
where $D^2 \equiv D_k D^k$.
These two constraints form a new pair of two second class constraints eliminating the scalar degree of freedom leading to a minimally modified theory of gravity.

The time conservation of $C_N$ and $C_R^{(2)}$ will fix the Lagrange parameter $u_N$ and $\lambda$
\begin{align}
    \dot C_N[\xi_1] & \approx \{ C_N[\xi_1] ,\int \md^3x \mathcal{H} \} + \{ C_N[\xi_1] , \int \md^3x \lambda C_R \} +\{ C_N[\xi_1], \int \md^3x u_N \pi_N \}, \\
    \dot C_R^{(2)}[\xi_2] & \approx \{ C_R^{(2)}[\xi_2] ,\int \md^3x \mathcal{H} \} + \{ C_R^{(2)}[\xi_2] , \int \md^3x \lambda C_R \} + \{ C_R^{(2)}[\xi_2], \int \md^3x u_N \pi_N \},
    \label{eq:second_consistency_relation}
\end{align}
where we have introduced the smeared constraints by using the test function $\xi(x)$
\begin{align}
    C_R[\xi] = \int \md^3x\, \xi C_R
\end{align}
and similar for the other constraints. Using that 
\begin{align}
    & \{ C_N[\xi_1], \int \md^3x \lambda C_R \} \nonumber \\ 
    & \approx \int \md^3y \frac{\delta C_N[\xi_1]}{\delta \pi^{ij}(y)}\sqrt{h} \left( - D_{(i} D_{j)} D^2 - D_{(i} R D_{j)} + R_{ij} D^2 + h_{ij} \left( D^4 - \frac{1}{2} D^k R D_k \right) \right) \lambda(y)
\end{align}
and similar for $\{ C_R^{(2)},\int\md^3x \lambda C_R \}$ we can see that it will lead to a spatial differential equation for $\lambda$. However, we can note that for a homogeneous and flat ansatz like the flat FLRW metric the second consistency relation \eqref{eq:second_consistency_relation} becomes trivial and $\lambda$ is not fixed. 

\subsection{Lagrangian formulation}
The new action can be obtained by an inverse Legendre transformation
\begin{align}
\label{eq:Toy_model_constraint}
    \mathcal{L}^\prime = \mathcal{L} +  \lambda C
\end{align}
where we have rescaled $\lambda \rightarrow - \lambda$.
In this case, the equation of motions (EOM) are given by
\begin{align}
   & \frac{\delta ( \int \md^4x\,  \mathcal{L} )}{\delta h_{ij} } + \sqrt{h} \left(  D_{(i} D_{j)} D^2 + D_{(i} R D_{j)} - R_{ij} D^2 - h_{ij} \left( D^4 - \frac{1}{2} D^k R D_k \right) \right) \lambda = 0, \nonumber \\
    & D^2 R =0, \qquad 
    \frac{\delta (\int \md^4x\,  \mathcal{L})}{\delta N} =0, \qquad 
    \frac{\delta (\int\md^4x\,  \mathcal{L})}{\delta N^k} =0.
\end{align}
Taking the trace of the first equation we can solve it for $\lambda$ yielding
\begin{align}
   \left(  2 D^4 + R D^2 + \frac{1}{2} D^k R D_k \right) \lambda = \frac{h_{ij}}{ \sqrt{h}} \frac{\delta (\int \md^4x\, \mathcal{L})}{\delta h_{ij}}.
\end{align}
We obtain a spatial differential equation for $\lambda$. The traceless component of the first EOM leads to
\begin{align}
    \left( h^{i}_m h^j_n - \frac{1}{3} h^{ij} h_{mn} \right) \left( \frac{\delta (\int\md^4x\, \mathcal{L}) }{\delta h_{mn}} + \sqrt{h} \left(D_{(i} D_{j)} D^2 + D_{(i} R D_{j)} - R_{ij} D^2  \right) \lambda \right)  =0.
\end{align}
Note that the EOM are invariant under a shift $\lambda(t,x^k) \rightarrow \lambda (t,x^k) + \lambda_0(t)$. As a consistency check we can count the number of degrees of freedom. The trace component of the metric EOM fixes $\lambda$, the Hamiltonian constraint and the momentum constraint fix $N$ and $N^k$ and finally the constraint $C$ fixes the trace of the metric. Using the remaining spatially gauge invariance we can fix 3 further components of the metric so that we are left with two traceless components of the spatial metric, $h_{ij}$.

\subsection{FLRW background}
Let us now apply this new theory to cosmology. At the FLRW background
\begin{align}
    \md s^2 = - N^2 \md t^2 + a^2 \md x^i \md x^j \delta_{ij}
\end{align}
we can see that both of the constraints vanish identically. In particular, the action at the minisuperspace is equivalent to the original K-essence model
\begin{align}
    \mathcal{L}^\prime_{\mathrm{FLRW}} = a^3 N \left( - 3 \left( \frac{\dot a}{aN} \right)^2 + P \right).
\end{align}
Consequently the Hamiltonian is given via
\begin{align}
    \mathcal{H}_{\mathrm{FLRW}} =  a^3 N \left( - \frac{p_a^2}{12 a^4} - P \right).
\end{align}
The lapse function is non-dynamical so that $\pi_N$ and the Hamiltonian constraint $\mathcal{H}_0$
\begin{align}
    \mathcal{H}_0 = a^3 \left( - \frac{p_a^2}{12 a^4} - P + P_X \frac{\dot \varphi^2}{N^2} \right)
\end{align}
form a pair of two second class constraints resulting in one dynamical degree of freedom. In contrast to other MMG models there is a dynamical scalar degree of freedom at the FLRW background. The introduction of the two constraints does not impact the background evolution since the two constraint vanish identically. As a minor consistency check in the appendix \ref{app:Non_trivial_background} it is shown that one can recover the flat FLRW solutions by starting from a generic stationary spherical symmetric background metric in which case the constraints are not trivial identities by imposing proper boundary conditions at spatial infinity. On the other hand, at the linear level the constraint $C$ will lead to 
\begin{align}
    \delta C = a \partial^2 \delta R  \approx 0 
\end{align}
which eliminates the scalar degree of freedom at the linear level.  

Note, that one can obtain MMG models with a dynamical FLRW background even without the usage of auxiliary constraints. However, the structure of these theories is very different requiring for instance a trivial Hamiltonian constraint at the FLRW background leading to pathological models (see appendix \ref{app:Dynamical_FLRW} for more details).

\subsection{Gravitational waves}
\label{subsec:Gravitational_waves}
To study the impact of the constraint on the EOM let us consider the consequences for the gravitational waves around a generic background 
\begin{align}
    h_{ij} = \bar h_{ij} + \delta h_{ij}, \qquad N=\bar N +\delta N, \qquad N^k = \bar N^k + \delta N^k.
\end{align}
We use our gauge symmetry to fix $\bar D_k \delta h^{k}_j=0$.
For simplicity let us focus only on the terms with the highest number of derivatives acting on $\delta h_{ij}$ which is similar to the geometrical optics approach in general relativity \cite{PhysRev.166.1263,PhysRev.166.1272}
\begin{align}
    \delta h_{ij} = A_{ij}  \exp(\theta/\epsilon)
\end{align}
with $\epsilon \ll 1$.
From the constraint EOM we obtain at leading order
\begin{align}
    \bar D^4 \delta h^k_k \simeq - \tilde{\bar R}^{ij} \bar D^2 \delta \tilde h_{ij}
\end{align}
where we have introduced the traceless tensor $\delta \tilde h^{ij} = \delta h^{ij} - h^{ij} \delta h^k_k/3$. Therefore, the trace of the metric perturbation is of order  $\delta h^k_k \simeq \mathcal{O}(\epsilon^2)$. In order to obtain the scaling relation for the shift and lapse we can use the Hamiltonian and momentum constraint. 

Perturbing the momentum constraint up to linear order yields
\begin{align}
    & - \left( \bar K^i_j \bar D_i - \bar K \bar D_j  \right) \delta N - \frac{1}{2}\bar D_j \bar D_m \delta N^m - \frac{1}{2} \bar D^2 \delta N_j + \frac{1}{2} \bar D^k \delta \dot h_{kj} \simeq 0.
\end{align}
Note that by using the gauge condition
 \begin{align}
    \bar D_j \delta \dot h^{jk} \simeq \epsilon^{-1}  \left( \dot A^{jk} \bar D_j \theta + \bar D_j (\dot \theta A^{jk})  \right).
\end{align}
On the other hand solving the Hamiltonian constraint up to leading order we obtain 
\begin{align}
    &\delta N \left( 2 \bar R + 2 P - \frac{2}{\bar N^2} P_X + \frac{4}{\bar N^4} P_{XX} \right) - \frac{4}{3} \bar K \bar D_b \delta N^b \nonumber \\
    &- \tilde{\bar K}^{ij} \left( \delta \dot h_{ij} - \bar D_i  \delta N_j - \bar D_j \delta N_i - \bar N^c \bar D_c \delta h_{ij} \right) \simeq \mathcal{O}(\epsilon^0).
\end{align}
Therefore, if  $\tilde{\bar K}_{ij}=0$, we can make the self-consistent ansatz $\delta N=\mathcal{O}(\epsilon^0)$ and $\delta N^k = \mathcal{O}(\epsilon)$. However, as long as the background has non-vanishing traceless components of the extrinsic curvature we need to require that either $\tilde{\bar K}^{ij} (\delta\dot h_{ij} - \bar N^c \bar D_c \delta h_{ij})=0$, which leads to non-dynamical gravitational waves, or that $\bar D_j \delta N^k$ and $\delta N$ scale as $\mathcal{O}(\epsilon^{-1})$. 

As a next step perturbing the trace of the EOM for $h_{ij}$ leads to
\begin{align}
    \bar D^4 \delta \lambda \simeq  \frac{1}{2} \delta \dot K  - \frac{1}{2} \bar N^k \bar D_k \delta K - \frac{1}{2} \bar D^2 \delta N  + \bar D^{i} \bar D^j \bar \lambda \bar D^2 \delta h_{ij}
\end{align}
For the traceless EOM the leading order is given by
\begin{align}
     & \left( \bar h^{i}_m \bar h^j_n - \frac{1}{3} \bar h^{ij} \bar h_{mn} \right) \Big[  - \frac{1}{2}\delta \dot K_{ij} + \frac{1}{2} \bar N^k \bar D_k \delta K_{ij} -  \frac{1}{2} \bar N  \delta R_{ij}    + \frac{1}{2} \bar D^2 \bar \lambda \bar D^2 \delta h_{ij} \nonumber \\
     & + \frac{\bar D_{(i} \bar D_{j)}}{\bar D^2} \left( \frac{1}{2} \delta \dot K  - \frac{1}{2} \bar N^k \bar D_k \delta K - \frac{1}{2} \bar D^2 \delta N   \right) \Big]  \simeq 0.
\end{align}
For $\tilde{\bar K}_{ij}\neq 0$, in which case $\delta N$ and $\bar D_k \delta N^j$ scale as $\mathcal{O}(\epsilon^{-1})$, we obtain $\delta \dot K=\mathcal{O}(\epsilon^{-2})$. Therefore, the dispersion relation for the gravitational waves is, in general, quite cumbersome. We need to solve the Hamiltonian and momentum constraint explicitly. It might be that the dispersion relation gets non-local contributions. 

On the other hand, if $\bar K_{ij}= \frac{1}{3} \bar K \bar h_{ij}$ we recover for the traceless spatial components the standard EOM for gravitational waves on a generic background up to a modified propagation speed which depends on the background value $\bar D^2 \bar \lambda$
\begin{align}
    & - \frac{1}{2 \bar N} \left( \delta \tilde{\ddot h}_{ij} - 2 \bar N^k \bar D_k \delta \tilde{\dot h}_{ij} \right)  - \frac{1}{2 \bar N} \bar N^k  \bar N^m \bar D_m \bar D_k \delta \tilde h_{ij} + \frac{1}{2}  \left( \bar N + \bar D^2 \bar \lambda \right) \bar D^2 \delta \tilde h_{ij} \nonumber \\
    & \simeq \frac{\bar N}{2} \bar g^{\alpha\beta} \bar \nabla_\alpha \bar \nabla_\beta \delta \tilde h_{ij} + \frac{1}{2} \bar D^2 \bar \lambda \bar D^2 \delta \tilde h_{ij} =0
\end{align}
where $\bar g^{\alpha\beta}=h^{\alpha\beta}-n^\alpha n^\beta$ and $n^\alpha=1/N(1,-N^k)$. 

Note, that the result is highly dependent on the form of the constraint. Consider for instance the constraint
\begin{align}
    \hat C_R = \sqrt{h} \left( R + \frac{R_{ij} R^{ij}}{\Lambda^2}\right).
\end{align}
In this case up-to-leading order the constraint yields
\begin{align}
    -\left(  D^2  + \frac{R}{3 \Lambda^2} D^2 + \frac{R^{ij}}{\Lambda^2} D_i D_j  \right) \delta h_c^c \simeq  \frac{\tilde R^{ij}}{\Lambda^2} D^2 \delta \tilde h_{ij}.
\end{align}
Therefore, the spatial metric is not anymore traceless up-to-leading order but instead the trace component is of the same order. In general, the EOM will lead to a higher order dispersion relation.

\section{Minimally modified gravity models}
\label{sec:Hamiltonian_construction}

\subsection{Construction with auxiliary Lagrange multiplier}

Let us shortly recap the framework to construct minimally modified gravity models with auxiliary Lagrange multiplier following \cite{Yao:2020tur}. 

Let us start with a generic Hamiltonian with a spatial diffeomorphism invariance
\begin{align}
    H_T = H_p + \int \md^3x\, N^k \mathcal{H}_k + u^i \pi_i
\end{align}
where again $\mathcal{H}_k$ is the standard momentum constraint which generates the spatial transformation.
On the other hand, $H_p$ can be expressed as
\begin{align}
    H_p = \int \md^3 x\, \mathcal{H}(N,h_{ij},\pi^{ij},D_k,t) + u_N \pi_N.
\end{align}
This class of models has in general six first class constraints, $\mathcal{H}_k$ and $\pi^k$, and two second class constraints $\pi_N$ and the Hamiltonian constraint $\mathcal{H}_0$. 

By imposing additional constraints at the Hamiltonian level we can remove the scalar degree of freedom. Following \cite{Yao:2020tur} and similar to the toy model in the previous section we could impose a new primary constraint at the Hamiltonian via
\begin{align}
    H_p^\prime = H_p +\int \md^3x\, \lambda C.
\end{align}
Further, we have to require that the consistency relation $\dot C \approx 0$ does not fix the Lagrange multiplier but instead leads to a secondary constraint which requires that $C$ commutes with itself and with $\pi_N$,
\begin{align}
    C= C(h_{ij},\pi^{ij},D_k,t)\quad \mathrm{with} \quad \{ C(x), C(y)\}\approx 0.
    \label{eq:Secondary_constraint}
\end{align}
Having a pair of second class constraint (a primary and a secondary one) the scalar degree of freedom is killed leading to minimally modified gravity theory with two tensor degrees of freedom. 

Alternatively, we could directly impose two primary constraints at the Hamiltonian \cite{Ganz:2021hmp}
\begin{align}
\label{eq:Two_constraints}
    H_p^\prime = H_p + \int \md^3x\,  \lambda_1 C_1 + \lambda_2 C_2
\end{align}
Having already a pair of second class constraints the only conditions on the form of $C_1$ and $C_2$ are that they commute with $\pi_N$ and are invariant under spatial diffeomorphism, i.e. $C_i=C_i(h_{ij},\pi^{ij},D_k,t)$.  Note, that the two approaches are not equivalent. In the latter case both constraints have to independent of the lapse function in order to commute with $\pi_N$ while the secondary constraint in the first approach will, in general, depend on the lapse function. On the other side in the second approach the two imposed constraints depend on the momentum of the metric. Even if the secondary constraint in \eqref{eq:Secondary_constraint} does not depend on the lapse function we would need to ensure that the two constraints in \eqref{eq:Two_constraints} can be expressed in such a way that one of them does not depend on the momentum in order to obtain an equivalent theory.

While the pair of second class constraints (either two primary or one primary and one secondary one) removes the scalar degree of freedom at the full non-linear level, our aim is it to keep the FLRW background dynamics of the original model without the imposed constraints. This enforces that the new constraints vanish identically on the FLRW background
\begin{align}
    C\vert_{\mathrm{FLRW}}=0.
\end{align}
In that case, similar to the discussed toy model in the previous section, at the FLRW background we end up with the same Hamiltonian as in the original model leading to one dynamical degree of freedom due to the broken full diffeomorphism invariance. 
\medskip

Note, that in \cite{Yao:2020tur} the authors also discussed another case where they only impose one new second class constraint but instead require that $\pi_N $ remains a first class constraint. This, however, restricts the form of the Hamiltonian to
\begin{align}
    \mathcal{H} =& \mathcal{V}(h_{ij},\pi^{ij},D_k,t) + N \mathcal{H}_0(h_{ij},\pi^{ij},D_k,t).
\end{align}
At the FLRW background this kind of models, in general, does not have a dynamical degree of freedom. Even if the new constraint $C$ vanishes at the background level $C\vert_{\mathrm{FLRW}}=0$ due to the homogeneous background $\mathcal{H}_0$ will commute with itself and, therefore, at the background level we will obtain two first class constraints as in standard GR resulting in a FLRW background without any dynamics. In order to obtain a dynamical FLRW background the Hamiltonian constraint itself has to become trivial at the background leading to a pathological behavior. This is similar to the toy model discussed in appendix \ref{app:Dynamical_FLRW}.

\if{}
\subsubsection*{1st class}
We add an additional primary constraint by hand to the Hamiltonian via 
\begin{align}
    H_p^\prime = H_p + \int \md^3x\, \nu C
\end{align}
Further, let us require that $\pi_N$ remains a first class constraint. In this case it is possible to solve the form of the Hamiltonian and the constraint analytically leading to \cite{Yao:2020tur}
\begin{align}
    \mathcal{H} =& \mathcal{V}(h_{ij},\pi^{ij},D_k,t) + N \mathcal{H}_0(h_{ij},\pi^{ij},D_k,t), \\
    C=& C(h_{ij},\pi^{ij},D_k,t),
\end{align}
where the form of the Hamiltonian constraint $\mathcal{H}_0$ and $C$ are arbitrary.

\fi

\subsection{Breakdown of the perturbation theory}
\label{subsec:Breakdown}
In the previous subsection we have constructed minimally modified gravity models with just two tensor degrees of freedom while keeping the background dynamics at the FLRW background by requiring that the new constraints vanish on the FLRW background. There is, however, one caveat. While the new constraints should vanish on the FLRW background they should be present at the linear level.

To demonstrate it let us consider one simple example where the primary constraint is given by $C= \sqrt{h} R_{ij} R^{ij}$. For the Hamiltonian we use again K-essence. The secondary constraint can then be obtained due to the consistency condition
\begin{align}
    \frac{\md C}{\md t} = \{C, H_T\} \approx 0.
\end{align}
The specific form is quite cumbersome but it is straightforward to see that the constraint will again be trivially fulfilled at the FLRW background. 
\if{}
\begin{align}
    C_{2} \simeq & -8 N R_{ij} R^{jk}  \pi_k^i + 2 R \pi D^2 N + 2 \pi N D^2 R + 2 N D_k R D^k N - 8 \pi^{ij} D_k N D^k R_{ij} \nonumber \\
    & + 4 \sqrt{h}  D^k \left(\frac{\pi}{\sqrt{h}} \right) D_k \left( R N \right) + 4 \pi^{ij} D_i R D_j N - 4 \sqrt{h} R^{ij} D_i N D_j \left( \frac{\pi}{\sqrt{h}} \right) - 2 \sqrt{h} N R^{ij} D_i D_j \left( \frac{\pi}{\sqrt{h}} \right) \nonumber \\
    & - N D^2 R \pi + 2 \sqrt{h} N R D^2 \left( \frac{\pi}{\sqrt{h}}\right) + 8 R^{ij} D_i D_k N \pi_j^k - 2 R^{ij} \pi D_i D_j N + 8 \pi^{ij} D^k N D_j R_{ki} \nonumber \\
    & - 4 R^{ij} \pi_{ij} D^2 N - 8 \sqrt{h} R^{ij} D^kN \left( D_k \left( \frac{\pi_{ij}}{\sqrt{h}}\right) - D_j \left( \frac{\pi_{ik}}{\sqrt{h}} \right) \right) + 8 N \pi^{ij} D_k D_j R_i^k- 4N \pi^{ij} D^2 R_{ij} \nonumber \\
    & - 4 \sqrt{h} N R^{ij} D^2\left(\frac{\pi_{ij}}{\sqrt{h}} \right) + 8 \sqrt{h} N D_k R_{ij} D^j \left( \frac{\pi^{ik}}{\sqrt{h}} \right) - 8 \sqrt{h} N D_k R_{ij} D^k\left( \frac{\pi^{ij}}{\sqrt{h}} \right).
\end{align}
Obviously the new constraints vanish on the FLRW background. 
\fi 
Since $C$ does not depend on the momentum the Legendre transformation is given by
\begin{align}
    \mathcal{L}^\prime = \mathcal{L} +  \lambda C.
\end{align}
However, considering the linear perturbation of the constraint $C$ we can see that the constraint is still trivially fulfilled
\begin{align}
    \delta C = 0.
\end{align}
Therefore, at the linear level there is no additional constraint and we end up with one scalar degree of freedom which is in strong contradiction with the full non-linear theory. Note, that it does not imply that the full non-linear theory is inconsistent. Instead, it means that we cannot trust perturbation theory around this given background. This is similar to the discussion of strong coupling where the linear scalar degree of freedom is absent at the linear level but returns at higher order. 

Therefore, in order to have a consistent perturbation theory around the FLRW we need to ensure that the constraints are not trivially fulfilled. This limits the possible number of operators. If we are only interested in the linear order we can expand the constraint in terms which vanish on the background resulting in 
\begin{align}
 c_1(t,\pi) R, \quad c_2(t,\pi) D_k D^k R, \quad c_3(t,\pi) D_k D^k \pi, ...
\end{align}
where the dots signal higher order of spatial derivatives. Each constraint can be written as a linear combination of the aforementioned terms and operators which vanish at the linear order. Note, that terms like $ D_j \pi^{ij}$ and $ D_j R^{ij}$ do not yield new independent operators due to the momentum constraint and the Bianchi identity. 
\medskip

Last, let us note that this also implies that a common assumption that the linear perturbation theory can be used to derive a lower bound on the number of degrees of freedom at the full non-linear level is in general not correct. As in the previous toy model the theory could contain constraints which vanish at the linear order for a given background leading to an inconsistent perturbation theory which overestimates the number of degrees of freedom (see also appendix \ref{app:Dynamical_FLRW} for an example without the presence of auxiliary constraints). 

\section{Effective field theory}
\label{sec:Effective_Field}

Let us use the derived framework to discuss phenomenological consequences at the linear order around the FLRW background for this kind of models. Our analysis will be split into two parts. First, we will impose one primary constraint, which does not depend on the momentum and generates a secondary constraint. In the second case we directly impose two primary constraints. Note, that we do not include cases where the single primary constraint also depends on the momentum since the analysis is much more involved due to the requirements on the form of the primary constraint and is beyond the scope of this paper. In order to have a dynamical scalar degree of freedom at the linear order we will consider pure dust for the matter sector.

\subsection{No momentum dependency}
We impose one primary constraint which does not depend on the momentum
\begin{align}
    H_p^\prime = \int \md^3x\, \mathcal{H} + u_N \pi_N + \lambda C
\end{align}
with
\begin{align}
    C = C(h_{ij},D_k,t).    
\end{align}
As discussed before, the Legendre transformation becomes trivial in that case 
\begin{align}
    \mathcal{L}^\prime = \mathcal{L} +  N \lambda C
\end{align}
where we have rescaled the Lagrange multiplier $\lambda \rightarrow - N \lambda$. Further, $\mathcal{L}$ is the Lagrangian associated to the Hamiltonian $\mathcal{H}$ without the presence of the constraint. 

In order to keep the formalism very general we will use the effective field formalism of dark energy for the Lagrangian \cite{Gleyzes:2013ooa}
\begin{align}
   \mathcal{L} = \sqrt{h} N L(K,S,R,Y,Z)
\end{align}
where $S= K_{ij} K^{ij}$, $Z=R_{ij} R^{ij}$ and $Y= K_{ij} R^{ij}$. Further, we will add pure dust described by the Schulz-Sorkin action (see appendix \ref{sec:Schultz-Sorkin}).

As discussed in section \ref{subsec:Breakdown}, in order to have a consistent perturbation theory around the FLRW background we can parametrize the constraint as 
\begin{align}
    C=  \sqrt{h} \sum c_n(t) (D_k D^k)^n R + g(h_{ij},D_k,t)
\end{align}
where $g$ is a free function which vanish at the background and linear order. 
For the metric perturbation we use the standard convention of the effective field formalism of dark energy
\begin{align}
    N= 1 + \delta N, \quad N^k= \delta^{ki} \partial_i \psi, \quad h_{ij}= a^2 e^{2\xi} \left( \delta_{ij} + \gamma_{ij} + \frac{1}{2} \gamma_{ik} \gamma^{k}_j + ... \right).
\end{align}
Similar we perturb the Lagrange multiplier as $\lambda = \lambda_0(t) + \delta \lambda$.
At linear order the constraint will result in a spatial differential equation for $\xi$
\begin{align}
    \delta C = - 4 a^3  \sum_n c_n \frac{\partial^{2n+2}}{a^{2n+2}}  \xi.
\end{align}
For simplicity, we consider boundary conditions so that $\xi=0$.
\medskip

We can split the final result into two different classes:
\begin{itemize}
    \item non propagating solution $\mathcal{A} + 2 L_S=0$
    \item propagating solution
\end{itemize}
Note, that all models inside the (beyond) Horndeski class \cite{Gleyzes:2014dya} belong to the first case.

\subsubsection*{Non-propagating solution}
In that case we can integrate out the perturbations of the non-dynamcial shift and lapse as
\begin{align}
    \delta N =& \frac{\rho_m}{\mathcal{B} + 4 H L_S}& , \\
    \frac{k^2}{a^2} \psi =& \frac{ (3 H \mathcal{B} - 2 L_N - L_{NN}  ) \rho_m v_m + \rho_m (\mathcal{B} + 4 H L_S) \delta_m } {(\mathcal{B} + 4 H L_S)^2}
\end{align}
where $L_X \equiv \partial_X L$ and
\begin{align}
    \mathcal{B} =&  2 H L_{SN} + L_{KN}, \\
    \mathcal{A} =& 4 H^2 L_{SS} + 4 H L_{SK} + L_{KK}.
\end{align}
Using further the EOM of $v_m$ 
\begin{align}
    v_m= \frac{a^2 (\mathcal{B} + 4 H L_S) ((\mathcal{B}+ 4 H L_S) \dot \delta_m + \rho_m \delta_m ) }{-k^2 (\mathcal{B} + 4 H L_S)^2 + a^2 ( 3 \mathcal{B}^2 \dot H + 24 \mathcal{B} L_S H \dot H + (2 L_N + L_{NN} ) \rho_m + 12 H^2 L_S (4 \dot H L_S + \rho_m)}
\end{align}
we finally obtain at the small scale limit $x= k/(aH) \gg 1$
\begin{align}
    \delta S= \int \md^3 k\md t\, \frac{a^5 \rho_m}{2 k^2} \Big[ \dot \delta_m^2 + \Big( \frac{ \rho_m^2}{( \mathcal{B} + 4 H L_S)^2} - \frac{1}{ a^5 \rho_m} \frac{\md }{\md t} \Big( \frac{a^5 \rho_m^2}{\mathcal{B} + 4 H L_S} \Big) \Big) \delta_m^2 \Big].
\end{align}
There is one dynamical non-progagating degree of freedom. The absence of ghost instabilities requires $\rho_m >0$ which is fulfilled for any canonical matter fluid. Further, up to the leading order in the small scale limit the EOM is given by
\begin{align}
    \ddot \delta_m + 2 H \dot \delta_m - \frac{1}{2} \rho_m G_{\mathrm{eff}} \delta_m=0
\end{align}
where
\begin{align}
    G_{\mathrm{eff}} =  2 \left( \frac{\rho_m}{(\mathcal{B} + 4 H L_S)^2}  - \frac{1}{a^5 \rho_m^2 } \frac{\md }{\md t} \Big( \frac{a^5 \rho_m^2}{ (\mathcal{B} + 4 H L_S)} \Big) \right).
\end{align}
The effective gravitational constant is changed. Note, that the expression is exact up to the leading order in the small scale approximation and does not require the quasistatic approximation since there is only one dynamical scalar degree of freedom.

\subsubsection*{Propagating solution}
In the second case we can again integrate out $\delta N$, $\psi$ and $v_m$
\begin{align}
    \delta N =& \frac{\rho_m (\mathcal{B} + 4 H L_S) }{\mathcal{D}} v_m  - \frac{\rho_m (\mathcal{A} + 2 L_S)}{\mathcal{D}} \delta_m , \\
   \frac{k^2}{a^2} \psi =& - \frac{\rho_m (2 L_N + L_{NN} - 3 H \mathcal{B} )}{\mathcal{D}} v_m - \frac{\rho_m (3 H \mathcal{A} + 2 H L_S - \mathcal{B})}{\mathcal{D}} \delta_m , \\
    v_m =& \frac{a^2 \mathcal{D} \dot \delta_m + a^2 (\mathcal{B} + 4 H L_S)\rho_m \delta_m }{- \mathcal{D} k^2 + 3 a^2 \dot H \mathcal{D} + a^2 \rho_m (2 L_N + L_{NN} + 12 H^2 L_S ) },
\end{align}
where
\begin{align}
    \mathcal{D}= \mathcal{B}^2 + 8 \mathcal{B} H L_S - 2 L_S ( 2 L_N + L_{NN} + 4 H^2 L_S ) - \mathcal{A} (2 L_N + L_{NN} + 12 H^2 L_S )
\end{align}
so that at the small scale limit $x= k/(aH) \gg 1$
\begin{align}
    \delta S= \int \md^3k \md t\, a^3 \Big[ & \frac{a^2 \rho_m}{2k^2} \dot \delta_m^2  + \Big( \frac{ \rho_m^2 (\mathcal{A} + 2 L_S) }{2 \mathcal{D}}  + \frac{a^2 \rho_m^3 ( \mathcal{B} + 4 H L_S)^2}{2 k^2 \mathcal{D}^2} \nonumber \\
    & - \frac{1}{ a^3} \frac{\md}{\md t} \big( \frac{a^5 \rho_m^2 (\mathcal{B}+ 4 H L_S)}{2 k^2\mathcal{D}}\big) \Big) \delta_m^2\Big].
\end{align}
In that case, the dust does not behave anymore as standard dust but instead acquires a non-vanishing sound speed
\begin{align}
    c_s^2 = - \frac{ (\mathcal{A} + 2 L_S)}{\mathcal{D}} \rho_m.
\end{align}
In order to have stable linear perturbations without gradient or ghost instabilities we have to require that $(\mathcal{A} + 2 L_S)/\mathcal{D}<0$ and $\rho_m >0$. Further, due to the strict constraints on the sound speed of dust we can use it to put severe constraints on the parameters of the model. 
Note, that the results do not depend on the background value of the Lagrange multiplier $\lambda_0$.

\subsubsection*{Tensor modes}
\label{subsec:No_momentum_dependency}

While the scalar sector does not depend on the background value $\lambda_0$ and is quite insensitive to the form of the constraints this is, in general, not the case for the tensor sector. Consider for instance $C=\sqrt{h} R$. The constraint will lead to a modification of the propagation speed which explicitly depends on $\lambda_0$. However, since the constraint is trivially fulfilled at the FLRW background, $\lambda_0$ is not fixed by the background EOM. Instead $\lambda_0$ is only fixed indirectly if we consider the full non-linear level.

On the other hand, in \ref{subsec:Gravitational_waves} we have discussed that for constraints like $C= \sqrt{h} D^2 R$ the EOM are invariant under a shift of $\lambda(t,x^k) \rightarrow \lambda(t,x^k) + \lambda_0(t)$. Therefore, we can set $\lambda_0=0$ without loss of generality similar to the discussion in \cite{Aoki:2020lig,Aoki:2020ila}. Indeed, we can check that in this case the EOM for the tensor sector will not be impacted by the constraint. 

In general, in order to avoid the ambiguity related to $\lambda_0$ we could restrict ourselves to constraints of the form $C=\sqrt{h} D_k C^k$ which ensures that the EOM do not depend on $\lambda_0$ and we recover at linear order the same result for the tensor sector as in the original model prior to the constraint.
\begin{align}
    \delta S = \frac{1}{4} \int \md^3k \md t\, a^3  \left( L_S \dot \gamma_{ij}^2 - \mathcal{E} \frac{k^2}{a^2} \gamma_{ij}^2  + (L_Z + \lambda C_Z) \frac{k^4}{a^4} \gamma_{ij}^2 \right)
\end{align}
where
\begin{align}
    \mathcal{E} = L_R  + \frac{1}{2 a^3} \frac{\md }{\md t} \left( a^3 L_Y \right).
\end{align}
Therefore, the sound speed of the tensor modes is given by
\begin{align}
    c_T^2 = \frac{\mathcal{E}}{L_S}.
\end{align}
Using the constraints from GW170817 \cite{PhysRevLett.119.161101,Abbott_2017} we can put severe constraints on the parameters of the model. Further, in order to avoid a higher order dispersion relation for the tensor modes we would need to set $L_Z =0$.

\subsection{Momentum dependency}

As a next step, let us discuss the case where we directly introduce two constraints which can in general depend on the momentum of the spatial metric
\begin{align}
    H_p^\prime= \int \md^3 x\, \mathcal{H} + u_N \pi_N + \lambda_1 C_1 + \lambda_2 C_2.
\end{align}
In order to have a dynamical FLRW background we have again to assume that the constraints are trivial identities on the background. Therefore, we can expand them as
\begin{align}
    C_1 =  \sqrt{h} \sum_k a_{1k}(t,\pi) (D_m D^m)^k R + \sum_k b_{1k}(t,\pi)   (D_m D^m)^k \pi + g_1(\pi^{ij},h_{ij},D_k), \\
    C_2 =  \sqrt{h} \sum_k a_{2k}(t,\pi) (D_m D^m)^k R + \sum_k b_{2k}(t,\pi)   (D_m D^m)^k \pi + g_2(\pi^{ij},h_{ij},D_k),
\end{align}
where $g_1$ and $g_2$ are arbitrary functions which vanish up to the quadratic order. Further, in order to have two tensor degrees of freedom at the full non-linear level the Dirac matrix has to be invertible. For simplicity, we assume that the two constraints do not commute with each other.

Perturbing both constraints at linear order we obtain for the scalar perturbations the following structure.
\begin{align}
    A(t,\partial,\pi,a)  \delta h_{ij} h^{ij} + B(t,\partial,\pi,a) \delta \pi=0.
\end{align}
These two constraints are in general differential equation. By imposing proper boundary conditions and requiring that the two constraints do not commute with each other we can set $\delta\pi=0=\delta h_{ij} h^{ij} $. This result is quite generic and does not depend on the specific form of the constraints. Therefore, in the following we will consider for simplicity 
\begin{align}
    C_1 = \sqrt{h} D_k D^k \left( \frac{\pi}{\sqrt{h}}\right), \qquad C_2 = \sqrt{h} D_k D^k R.
\end{align}
Further, to have an explicit expression for the Legendre transformation we will use the following ansatz for the Hamiltonian $\mathcal{H}$
\begin{align}
    \mathcal{H} = N d_0(N) \pi + N d_1(N) \pi_{ij} \pi^{ij} + N d_2(N) \pi^2 - N \sqrt{h} f(N,h_{ij},D_k)
\end{align}
where we assume that $f$ does not depend on $D_k N$ or higher order derivatives.
Performing the Legendre transformation we obtain the Lagrangian
\begin{align}
    \mathcal{L}_{tot} \equiv & \mathcal{L} + \lambda_2 C_2 \nonumber \\=& \sqrt{h} N \big[ \frac{1}{d_1} \mathcal{K}_{ij} \mathcal{K}^{ij} - \frac{d_2}{d_1 (d_1+3d_2)} \mathcal{K}^2 + f \big] + \lambda_2 C_2
\end{align}
where we have again rescaled $\lambda_2 \rightarrow - \lambda_2$ and
\begin{align}
    \mathcal{K}_{ij} = K_{ij} - \frac{1}{2} d_0 h_{ij} + \frac{1}{2N} h_{ij} D^2 \lambda_1.
\end{align}
As in the previous section the EOM for $\lambda_2$ leads to the known condition $C_2=\sqrt{h} D^2 R=0$ while taking the trace of the metric EOM we can solve $\lambda_2$ as
\begin{align}
   \left(  2 D^4 + R D^2 + \frac{1}{2} D^k R D_k \right) \lambda_2 = \frac{h_{ij}}{ \sqrt{h}} \frac{\delta (\int \md^4x\, \mathcal{L})}{\delta h_{ij}}.
\end{align}
On the other hand, the EOM for $\lambda_1$ leads to
\begin{align}
    D^2 \left( \frac{1}{d_1 + 3 d_2} \mathcal{K} \right) =0.
\end{align}
We can note, that in both cases the Lagrange parameters $\lambda_1$ and $\lambda_2$ are only solved up to a time dependent integration constant which is fixed by the boundary conditions of the spatial differential equations.

\subsubsection*{Scalar perturbations}
Let us now discuss the impact of the constraints on the cosmological scalar perturbations. As before, we will add a perfect fluid of pure dust to have a dynamical scalar degree of freedom at the linear level.

At linear order the curvature constraint $\delta C_2$ leads again to $\xi=0$ if we assume proper boundary conditions. Further, solving the EOM for $\delta \lambda_1$ we can integrate it out. Expanding the action in the small scale limit $x=k/(aH) \gg 1$ we obtain finally 
\begin{align}
    \delta_2 S = \int \md^3k \md t\, a^3 z^2 \big[ \dot \delta_m^2 - \left( c_s^2 \frac{k^2}{a^2} + M^2\right) \delta_m^2 \big]
\end{align}
where 
\begin{align}
    z^2 =& \frac{\rho_m a^2}{2 k^2}, \\
    c_s^2=& - \frac{4 \rho_m (d_1 + 3 d_2)^2}{b} 
\end{align}
with
\begin{align}
    b=& - 6 d_0 (d_1 + 3 d_2) (2 d_0^\prime + d_0^{\prime\prime} ) + (3 d_0^2+12 H^2) ( 2 d_1^\prime + 6 d_2^\prime + d_1^{\prime\prime} + 3 d_2^{\prime\prime} ) \nonumber \\
     & + 12 H (d_1 + 3 d_2)  (2 d_0^\prime + d_0^{\prime\prime} ) - 12 H d_0  ( 2 d_1^\prime + 6 d_2^\prime + d_1^{\prime\prime} + 3 d_2^{\prime\prime} ) \nonumber \\
     &+ 4 (d_1 + 3 d_2)^2 (2 f^\prime + f^{\prime\prime}) 
\end{align}
where the $^\prime$ denote derivatives with respect to the lapse function. 
The explicit expression of $M$ is quite involved and not really helpful for our purposes. We can note that the dust will acquire a non-vanishing sound speed which could be used to constrain the parameters of the model. Indeed, the sound speed only vanishes if $d_1=-3d_2$ which corresponds to a model which depends linearly on the trace of the momentum of the spatial metric.

\subsubsection*{Tensor perturbations}
Due to the form of the constraints $C_1$ and $C_2$ as total spatial derivatives
they do not impact the tensor sector at linear order but instead we recover the same EOM as for the original model which will depend on the form of the free function $f$. If we for instance consider the case where $f=f(R,Z)$ we obtain
\begin{align}
    \delta S = \frac{1}{4} \int \md^3k \md t\, a^3  \left( \frac{1}{d_1} \dot \gamma_{ij}^2 - f_R \frac{k^2}{a^2} \gamma_{ij}^2  + f_Z \frac{k^4}{a^4} \gamma_{ij}^2 \right),
\end{align}
so that the propagation speed of the gravitational waves is given by
\begin{align}
    c_t^2 = f_R d_1.
\end{align}

\bigskip

\if{}
For the matter we consider a relativistic perfect fluid following ...
\begin{align}
    H_{\mathrm{matter}}=& \int \md^3 N \mathcal{H}^{\mathrm{matter}}_{0} + N^j \mathcal{H}^{\mathrm{matter}}_{j}, \\
    \mathcal{H}^{\mathrm{matter}}_{0}=& \frac{p_\phi^2}{\sqrt{h}\mu^{\alpha-2}}-\sqrt{h} \mu^{\alpha}, \\
    \mathcal{H}^{\mathrm{matter}}_{j} =& p_\phi \partial_j \phi
\end{align}
where $\mu$ is implicitly defined in terms of $p_\phi$ via
\begin{align}
     \mu^2 = \left( \frac{p_\phi}{\sqrt{h} \mu^{\alpha-2}}\right)^2 + h^{ij} \partial_i \phi \partial_j \phi.
\end{align}
The fluid is characterized by
\begin{align}
    p(\mu) = \frac{1}{\alpha} \mu^\alpha, \quad \alpha = \frac{1+\omega}{\omega}
\end{align}
where $\omega$ is the equation of state $p=\omega \rho$. 

In the following we will focus on the scalar part.
To be general we consider an effective framework for the Hamiltonian constraint similar to the effective field formalism in dark energy, namely
\begin{align}
    \mathcal{H}_g = \sqrt{h} H_g(N,\pi/\sqrt{h}, S_\pi, R,Z, Y_\pi,t)
\end{align}
where 
\begin{align}
    S_\pi= \frac{1}{h} \pi^{ij} \pi_{ij}, \quad
    Y_\pi =  \frac{1}{\sqrt{h}}\pi^{ij} R_{ij}
\end{align}
It is important to note that one cannot directly relate $S_\pi$ and $Y_\pi$ to $S$ and $Y$ from the previous section. Indeed considering just GR plus a term linear in $Y_\pi$ we obtain after performing the Legendre transformation the operators $Y$ and $Z=R_{ij}R^{ij}$ in the Lagrangian. 

Defining the perturbations as
\begin{align}
    \delta \pi^{ij} = \pi^{ij} - \frac{1}{3} p \delta^{ij}, \quad \delta h_{ij} = h_{ij} - a^2 \delta_{ij}
\end{align}
with $\{ a^2 , p \}=1$,
at the background level the Hamiltonian simplifies to 
\begin{align}
    \mathcal{H}_g = a^3 H_g ( N, p/a,1/3 p^2 a^{-2} ,0,0,t), \\
    \mathcal{H}_m = a^3 \frac{\alpha-1}{\alpha} \left( \frac{p_\phi}{a^3} \right)^{\frac{\alpha}{\alpha-1}}
\end{align}
Therefore, we can see that at the background the momentum $p_\phi$ is constant in time. In the limit $\alpha \gg 1$ it represent the total energy density $a^3 \rho_m$.

Going into Fourier space and following the convention in \cite{} we will parametrize the  perturbations as
\begin{align}
   \delta h_{ij} = \delta q_m A^m_{ij}, \qquad \delta \pi^{ij} = \delta \pi^n A_n^{ij}
\end{align}
where
\begin{align}
    & A^1_{ij} = \delta_{ij}, \quad A^2_{ij} = \left( \frac{k_i k_j}{k^2} - \frac{1}{3} \delta_{ij} \right), \quad A^3_{ij} = \frac{1}{\sqrt{2}} \left( \hat k_i \hat v_j + \hat k_j \hat v_i\right), \quad A^4_{ij}=  \frac{1}{\sqrt{2}} \left(\hat k_i \hat w_j + \hat k_j \hat w_i\right),\nonumber \\
   & A^5_{ij} = \frac{1}{\sqrt{2}} \left( \hat v_i \hat w_j + \hat v_j \hat w_i  \right), \quad A^6_{ij} = \frac{1}{\sqrt{2}} \left( \hat v_i \hat v_j - \hat w_i \hat w_j\right)
\end{align}
where $k^2= k_i k_j \delta^{ij}$, $\hat k^i = k^i/\vert \vec k \vert$ and $\hat v_i$ and $\hat w_j$ are the orthonormal vector to $\hat k_i$ spanning the whole space. Similar $A_n^{ij}$ is defined such a way that $A^m_{ij} A_n^{ij} = \delta^m_n$. In the following we will not consider the vector modes $\delta q_3$, $\delta q_4$, $\delta \pi^3$ and $\delta\pi^4$ since they are not relevant for our purpose. 

\subsubsection*{Scalar perturbations}

Let us first focus on the scalar modes. 
We introduce the gauge fixing condition $\delta q_2 =0$ which forms together with $\delta \mathcal{H}^g_{\vec k} = - i \mathcal{H}_j k^j/k^2 $ a pair of two second class constraints. 

Together with the gauge fixing conditions we have 6 second class constraints, $\delta C_1$, $\delta C_2$, $\delta \pi_N$, $\delta \mathcal{H}_0$, $\delta q_2$, $\delta \mathcal{H}_{\vec k}$. The constraints $C_1$ and $C_2$ leads to 
\begin{align}
    c_q(t,p a^2,k) \delta q_1 + c_\pi(t,p a^2 ,k) \delta \pi_1 =0
\end{align}
and similar for the other one. Requiring that the two constraints do not commute with each other leads to $c_q d_\pi - c_\pi d_q \neq 0$. The only valid solutions are then given by $\delta \pi_1 = \delta q_1 =0$. Consequently, it is convenient to re-express $\delta C_1$ and $\delta C_2$ as just $\delta C_q = \delta q_1 \approx 0$ and $\delta C_\pi= \delta \pi_1 \approx 0$.

The gravitational part of the first order Hamiltonian, $\mathcal{H}_0$, and the scalar part of the momentum constraint yields
\begin{align}
    \delta \mathcal{H}^g_0 =& a^3 \Big[ H_{g,NN} \delta N + H_{g,NR} 2 a^{-4} k^2 \left( \delta q_1 - \frac{1}{3} \delta q_2  \right) + H_{g,NS_\pi}  \frac{p}{3a^4}(2 a^2 \delta \pi^1 - p \delta q_1) \nonumber \\
    & + \frac{1}{2 a^3} H_{g,N\pi} \left( 2 a^2 \delta \pi^1 -  p \delta q_1  \right) + \frac{2}{3}  H_{g,NY_\pi} p a^{-5} k^2 \left( \delta q_1 - \frac{1}{3} \delta q_2 \right)  \Big], \\
    \delta \mathcal{H}^g_{\vec k} =& - i \delta \mathcal{H}_j \frac{k^j}{k^2} = a^2 \left( -\frac{2}{3} \delta \pi_1 - 2 \delta \pi_2 + \frac{a^{-2} \pi }{3} ( \delta q_1 - \frac{4}{3} \delta q_2) \right) 
\end{align}
Note, that the other components of the momentum constraint only contain vector modes. 

Similar for the matter part
\begin{align}
    \delta \mathcal{H}^m_0 =& \delta P, \\
    \delta \mathcal{H}^m_{\vec k} =& P \delta T
\end{align}

It is then straightforward to calculate the Dirac matrix and check that the matrix is indeed invertible. Using the Dirac brackets the physical Hamiltonian can be obtained by strongly solving the set of second class constraints and then inserting it into 
\begin{align}
    \delta H_{\mathrm{phys}} =( \delta_2 \mathcal{H} + \delta_2 \mathcal{H}_{m}) \vert_{C_A =0}
\end{align}
where $C_A$ stands for the complete set of the second class constraints. Using the constraints we obtain finally
\begin{align}
    \delta H_{\mathrm{phys}} = - \frac{1}{2 a^3 H_{g,NN}} \delta P^2 + \frac{k^2}{2 a^2} P \delta T^2 + \frac{3 P^2}{8 a^3} H_{g,S_\pi}  \delta T^2 
\end{align}
It is convenient to perform a canonical transformation to express the variables in physical quantities. 
The matter overdensity $\delta_m$ is given by
\begin{align}
    \delta_m = \frac{\delta \rho}{\rho} =  \frac{\delta P}{P} - \frac{3 \delta q_1}{2 a^2}
\end{align}
We perform the canonical transformation $(\delta P,\delta T) \rightarrow (\delta_m, \delta \tilde T) $ with 
\begin{align}
    \delta \tilde T= P \delta T
\end{align}
In the new canonical conjugate variables 
\begin{align}
    \delta H_{\mathrm{phys}} = &  - \frac{P^2}{2 a^3 H_{g,NN}} \delta_m^2 + \frac{k^2}{2 a^2 P} \delta \tilde T^2 + \frac{3 }{8 a^3} H_{g,S_\pi} \delta \tilde T^2  
\end{align}
To have stable linear perturbations we need that $H_{g,NN} <0$, which is, for instance, the case in Quintessence models. 
The EOM are given by
\begin{align}
\delta \dot{\tilde T} =& - \frac{P^2}{a^3 H_{g,NN}} \delta_m, \\
\dot \delta_m =& - \frac{k^2}{a^2 P} \delta \tilde T - \frac{3}{4 a^3} H_{g,S_\pi} \delta \tilde T
\end{align}
In the subhorizon limit we can express the decoupled EOM for $\delta_m$ as
\begin{align}
    & \ddot \delta_m + 2 H \dot \delta_m -\frac{\rho_m}{ H_{g,NN}}  \left( \frac{k^2}{a^2 } + \frac{3 }{4 } \rho_m H_{g,S_\pi}  \right)  \delta_m =0
\end{align}
We can see that the matter overdensity is propagating, i.e. the sound speed is non-vanishing
\begin{align}
    c_s^2 = - \frac{\rho_m}{H_{g,NN}}.
\end{align}
As mentioned before, this constrains $H_{g,NN} <0$. Further, the sound speed of dark matter is highly constrained by current measurements. This can be used to put severe constraints on the parameters of the model. For instance, standard Quintessence $H_{g,NN} \sim \dot \varphi^2$ leads to a highly inconsistent sound speed deep in the matter domination epoch. 

\subsubsection*{Tensor modes}

Let us now consider the tensor modes. Since the new constraints do not impact the tensor modes up to linear order around the FLRW background, the EOM are the same as for the original starting model. 
\begin{align}
    \delta H_{\mathrm{phys}} 
    = & a H_{g,S_\pi} \left( (\delta \pi^5)^2 + (\delta \pi^6)^2 \right) +  \left( \frac{4 p}{3 a} H_{g,S_\pi} + H_{g,\pi} + \frac{k^2 }{2a^2} H_{g,Y_\pi}\right) (\delta q_5 \delta \pi^5 + \delta q_6 \delta \pi^6) \nonumber \\
    & + \left( \frac{5 p^2}{18 a^3} H_{g,S_\pi} + \frac{p}{4a^2} H_{g,\pi} - \frac{1}{4 a} H_{g} - \frac{k^2}{4 a^3} H_{g,R} + \frac{p k^2 }{12 a^4} H_{g,Y_\pi} + \frac{k^4}{4a^5} H_{g,Z}  \right) ( \delta q_5^2 + \delta q_6^2) 
\end{align}
Let us first rescale the momentum and the generalized coordinate via
\begin{align}
    \delta \tilde \pi^{5,6} = \sqrt{2 H_{g,S_\pi}} a \delta \pi^{5,6}, \qquad \delta \tilde q_{5,6} = \frac{1}{\sqrt{2  H_{g,S_\pi}} a} \delta q_{5,6}
\end{align}
which leads to the extra term in the Hamiltonian
\begin{align}
    \delta H_{\mathrm{ext}} = - \frac{ \partial_t (a^2 H_{g,S_\pi})}{2 a^2 H_{g,S_\pi}}  \left( \delta \tilde \pi^5  \delta \tilde q_5 + \delta \tilde \pi^6  \delta \tilde q_6 \right)
\end{align}
where $\partial_t (a H_{g,S_\pi})$ should be understood as a function of the background variables $p$, $a$, $P$ and $T$ by using the background EOM. 
As a next step we diagonalize the kinetic term via the canonical transformation
\begin{align}
    \delta\Pi^{5,6} = & \delta \tilde \pi^{5,6} + B \delta \tilde q_{5,6} \nonumber \\
    =& \delta \tilde \pi^{5,6} + a \left( \frac{4 p}{3 a} H_{g,S_\pi} + H_{g,\pi} + \frac{k^2 }{2a^2} H_{g,Y_\pi} - \frac{ \partial_t (a^2 H_{g,S_\pi})}{2 a^2 H_{g,S_\pi}}  \right) \delta\tilde q_{5,6}, \\
    \delta Q_{5,6} =& \delta \tilde q_{5,6}
\end{align}
The Hamiltonian is now given by
\begin{align}
    \delta H_{\mathrm{phys}} = & \frac{1}{2 a} \left( (\delta \Pi^5)^2+ (\delta \Pi^{6})^2 \right) + \Big[ \left( 2 a^2 H_{g,S_\pi} \right) \Big( \frac{5 p^2}{18 a^3} H_{g,S_\pi} + \frac{p}{4a^2} H_{g,\pi} - \frac{1}{4 a} H_{g}  \nonumber \\
    & - \frac{k^2}{4 a^3} H_{g,R} + \frac{p k^2 }{12 a^4} H_{g,Y_\pi} + \frac{k^4}{4a^5} H_{g,Z}  \Big) - \frac{1}{2} \dot B - \frac{B^2}{2 a} \Big] (\delta Q_5^2 + \delta Q_6^2)
\end{align}
Note, that  $H_{g,Z} \neq 0$ or $H_{g,Y_\pi}\neq 0$ will lead to a higher order dispersion. This can be easily understood by remembering that $Y_\pi$ is related to $Y$ and $Z$ at the Lagrangian level. The sound speed is given by
\begin{align}
    c_s^2  =&  H_{g,S_\pi} \left( - H_{g,R} + \frac{p}{3 a} H_{g,Y_\pi} \right) - \frac{1}{a} \frac{\md}{\md t} \left( \frac{H_{g,Y_\pi}}{2 a} \right) \nonumber \\
    &-  \frac{H_{g,Y_\pi}}{a} \left( \frac{4p}{3a} H_{g,S_\pi} + H_{g,\pi} - \frac{\partial_t(a^2 H_{g,S_\pi})}{2 a^2 H_{g,S_\pi}}\right)
\end{align}
\fi

\if{} 
\subsubsection{Toy model}
As a consistency check let us consider one toy model. The original theory is given by standard Quintessence in the unitary gauge
\begin{align}
    \mathcal{H}_Q =  N \frac{2}{\sqrt{h}} (\pi^{ij}\pi_{ij}-\frac{1}{2} \pi^2) - \frac{1}{2}  \sqrt{h}  N R - \frac{\sqrt{h} \dot \varphi^2}{2 N} + \sqrt{h} N V(\varphi)
\end{align}
For the constraints we take
\begin{align}
    C_1 = \sqrt{h} R, \quad C_2 = D_k D^k \pi
\end{align}
which do not commute.  Last, for the matter we use 
\begin{align}
    L_m = \int \md^4x\, \sqrt{-g} \left(-\frac{1}{2} \partial_\mu \chi \partial^\mu \chi \right)^{\alpha/2}
\end{align}
Using the EOM for the metric we can express the momentum $\pi_{ij}$ in terms of the extrinsic curvature $K_{ij}$
\begin{align}
  \pi_{ij}=  \frac{\sqrt{h}}{2}  \left( K_{ij} - K h_{ij}\right) +  \frac{\sqrt{h}}{2 N} h_{ij} D_k D^k u_1
\end{align}
Performing the Legendre transformation the action is given by
\begin{align}
 S=& \int \md^4x\, \sqrt{h} N \Big( \frac{1}{2} K_{ij}K^{ij} - \frac{1}{2} K^2 + \frac{1}{2} R + \frac{1}{2 N^2} - V(t)  \Big)  \nonumber \\ 
 &+ \int \md^4x\, \sqrt{h} \Big( K D_k D^k u_1 -\frac{3}{4 N} D_k D^k u_1 D_j D^j u_1 + u_2 R  \Big) + S_m
\end{align}
While by construction the constraints vanish at zero order, at linear order we obtain
\begin{align}
    \delta C_1 = \partial^2 \delta K - K^{ij} \partial^2 \delta h_{ij} - \frac{3}{2} \partial^4 \delta u_1, \quad \delta C_2 =  \delta R
\end{align}
The first one fixes $\delta u_1$, while the second one sets $\xi =0$ by imposing proper boundary conditions. 

Plugging it back into the action we obtain for the new contribution
\begin{align}
    S_{\mathrm{constraint}} = \int \md^3 k \md t\, a^3 \frac{1}{3} \left( -\frac{k^2}{a^2} \psi + 3 H \delta N \right)^2
\end{align}
The contribution from the Quintessence yields
\begin{align}
    S_Q = \int \md^3k \md t\, a^3 \Big( (\frac{1}{2} \dot \varphi^2- 3 H^2)  \delta N^2 - 2 H \delta N \frac{k^2}{a^2} \psi    \Big)
\end{align}
Last, the matter part yields
\begin{align}
    S_m = \int \md^3 \md t\,a^3 \frac{P_X}{c_s^2} \Big( -  \frac{c_s^2 k^2}{2 a^2 } \delta \chi^2 - c_s^2 \dot \chi \delta \chi \frac{k^2}{a^2} \psi + X \delta N^2 -\dot \chi \delta N \delta \dot \chi + \frac{1}{2} \delta \dot \chi^2  \Big)
\end{align}
It is useful to express $\delta \chi$ in physical terms as
\begin{align}
    \delta_m = \frac{\rho_m + p_m}{\rho_m} \big( \frac{\delta \dot \chi}{\dot \chi} - \delta N \big) + 3\frac{\rho_m + p_m}{\rho_m} \xi
\end{align}
Similar to ... we can replace the variables which leads to 
\begin{align}
    S_m = \int \md^3 k\md t\,& a^3 \Big( \frac{a^2 \rho_m^2}{2 k^2(\rho_m + p_m)} \left( \dot \delta_m + \frac{k^2(\rho_m + p_m)}{a^2 \rho_m} \psi \right)^2 - \frac{\rho_m }{2(\rho_m + p_m)} \Big(\rho_m c_s^2 \nonumber \\
    & + \frac{3 a^2}{k^2} \big[ 5 H^2 (\rho_m c_s^2 - p_m) + \frac{\md }{\md t} \big( H (\rho_m c_s^2 -p_m\big)  \big] \Big) \delta_m^2  \nonumber \\
    & -\rho_m \delta N \delta_m + 3 H (\rho_m c_s^2 - p_m) \psi \delta_m \Big)
\end{align}
To simplify, we will only consider the two limits: stiff matter $c_s^2=1$, $\rho_m=p_m$ and dust $c_s^2=p_m=0$. Further, we will use the subhorizon approximation. In this case we obtain in the case of stiff matter
\begin{align}
    S= \int\md^3k \md t\,a^3  \frac{a^2 \rho_m}{4 k^2 } \Big[ \dot \delta_m^2 - \left( c_s^2 \frac{k^2}{a^2} + m^2 \right)\delta_m^2 \Big]
\end{align}
where 
\begin{align}
     c_s^2 =& 1- \frac{2 \rho_m }{24 H^2 -\dot \varphi^2}, \\
     m^2 =& - \frac{144 H^2 \rho_m^2 }{(-24 H^2 + \dot \varphi^2)^2} - \frac{1}{a^3 \rho_m} \frac{\md }{\md t} \left( \frac{12 a^5 H \rho_m^2 }{24 H^2 - \dot \varphi^2} \right)
\end{align}
and for dust we obtain
\begin{align}
     S= \int\md^3k \md t\,a^3  \frac{a^2 \rho_m}{2 k^2 } \Big[ \dot \delta_m^2 - \left( c_s^2 \frac{k^2}{a^2} + m^2 \right)\delta_m^2 \Big]
\end{align}
where
\begin{align}
    c_s^2 =& -\frac{\rho_m}{24 H^2 -\dot \varphi^2}, \\
    m^2 =& -\frac{36 H^2 \rho_m^2 }{(-24 H^2 + \dot \varphi^2)^2}  - \frac{1}{a^3 \rho_m} \frac{\md }{\md t} \left( \frac{12 a^5 H \rho_m^2 }{24 H^2 - \dot \varphi^2}\right)
\end{align}

\fi

\if{}
\section{K-essence minimally modified gravity}
Let us discuss the consequences of the constraints for K-essence models. 
\begin{align}
    H_T = \int \md^3x\, \mathcal{H} + N^k \mathcal{H}_k + u_N \pi_N + u_R C_R
\end{align}
where 
\begin{align}
    \mathcal{H} = N \mathcal{H}_{GR} - \sqrt{h} N P(\dot \varphi^2/N^2). 
\end{align}
We can use the remaining invariance under time-reparametrizations to set $\varphi=t$ so that the Hamiltonian is not explicitly time dependent. 
For simplicity, let us consider the constraint
\begin{align}
    C_R = \sqrt{h} R
\end{align}
The time conservation of $C_R$ leads to 
\begin{align}
    C_R^{(2)} \approx - 4 N \tilde R_{ij} \tilde \pi^{ij} + 4 \pi^{ij} D_i D_j N
\end{align}
On the other hand, the time conservation of $\pi_N$ yields
\begin{align}
    C_N = \mathcal{H}_{GR} - \sqrt{h} \left( P - 2  \frac{\dot \varphi^2}{N^2} P_X \right) \approx 0
\end{align}
The time conservation of $C_N$ and $C_R^{(2)}$ will fix the Lagrange parameter $u_N$ and $u_R$. 
In the following let us introduce the smeared constraints by using the test function $\xi(x)$
\begin{align}
    C_R[\xi] = \int \md^3x \xi C_R
\end{align}
and similar for the other constraints. 
The Dirac bracket can by expressed as
\begin{align}
    \Omega_{IJ} = \begin{pmatrix} 0 & 0 & \{ \pi_N[\xi_1], C_N[\xi_2] \} & \{ \pi_N[\xi_1], C_R^{(2)}[\xi_2]\}\\
    0 & 0 &  \{C_R[\xi_1], C_N[\xi_2] \} & \{ C_R[\xi_1] , C_R^{(2)}[\xi_2] \} \\
    - \{ \pi_N[\xi_1], C_N[\xi_2] \} & - \{C_R[\xi_1], C_N[\xi_2] \} & \{  C_N[\xi_1],  C_N[\xi_2] \} & \{  C_N[\xi_1], C_R^{(2)}[\xi_2]\} 
    \end{pmatrix}
\end{align}
In general the Lagrange parameter $u_R$ can be obtained by solving this system of constraints
\begin{align}
    \dot C_N[\xi_1] =& \{ C_N[\xi_1], H_T \} \approx \{ C_N[\xi_1] ,\int \md^3x \mathcal{H} \} + \{ C_N[\xi_1] , \int \md^3x u_R C_R \} +\{ C_N[\xi_1], \int \md^3x u_N \pi_N \}, \\
    \dot C_R^{(2)}[\xi_2] =& \{ C_R^{(2)}[\xi_2], H_T \} \approx \{ C_R^{(2)}[\xi_2] ,\int \md^3x \mathcal{H} \} + \{ C_R^{(2)}[\xi_2] , \int \md^3x u_R C_R \} + \{ C_R^{(2)}[\xi_2], \int \md^3x u_N \pi_N \}
\end{align}
It will lead to a spatial differential equation for $u_R$.
\begin{align}
     & \sqrt{h} \left( \frac{\delta C_N[\xi_1]}{\delta \pi^{ij}} + \frac{\delta C_R^{(2)[\xi_2]}}{\delta N}\left( - \frac{\delta C_N[\xi_1]}{\delta N}\right)^{-1} \frac{\delta C_R^{(2)}[\xi_2]}{\delta \pi^{ij}} \right)  \left[ \left( - D^i D^j  + h^{ij} D^2 \right) u_R + R^{ij} u_R \right]       \nonumber \\
     &   \approx 
     -  \{ C_R^{(2)}[\xi_2] ,\int \md^3x \mathcal{H} \} + \frac{\delta C_R^{(2)[\xi_2]}}{\delta N} \left(  \frac{\delta C_N[\xi_1]}{\delta N}\right)^{-1}  \{ C_N[\xi_1] ,\int \md^3x \mathcal{H} \}
\end{align}

\subsection{Lagrangian formalism}
Let us alternatively directly start from the Lagrangian formalism. 
In this case, the EOM are given by
\begin{align}
    \frac{\delta ( \sqrt{h} N \mathcal{L} )}{\delta h_{ij} } + \sqrt{h} \left( h^{kj} h^{li} - h^{kj} h^{ij} \right) D_k D_l \lambda  - \sqrt{h} R^{ij} \lambda = 0, \\
    \sqrt{h} R =0, \qquad 
    \frac{\delta (\sqrt{h} N \mathcal{L})}{\delta N} =0, \qquad 
    \frac{\delta (\sqrt{h} N \mathcal{L})}{\delta N^k} =0
\end{align}
Taking the trace of the first equation we can solve it for $\lambda$ yielding
\begin{align}
    D_k D^k \lambda = \frac{h_{ij}}{2 \sqrt{h}} \frac{\delta (\sqrt{h} N \mathcal{L})}{\delta h_{ij}}
\end{align}
Therefore, we obtain a Poisson equation for $\lambda$ so that the solution is given by
\begin{align}
    \lambda = \lambda_0(t) + \lambda_{par}(t,x^k)
\end{align}
Note, that for the homogeneous solution we have assumed that the manifold has no boundary so that from $D^2 \lambda_{hom}=0$ follows $\lambda_{hom}=\lambda_0(t)$, where $\lambda_0(t)$ is the integration constant.
Using the solution for $\lambda$ we can obtain the traceless component of the first EOM
\begin{align}
    \left( h^{i}_m h^j_n - \frac{1}{3} h^{ij} h_{mn} \right) \left( \frac{\delta (\sqrt{h} N \mathcal{L}) }{\delta h_{mn}} + \sqrt{h} { D^n D^m} \lambda \right) - \sqrt{h} \tilde R^{ij} \lambda =0.
\end{align}
This equation depends on the homogeneous part of $\lambda$ if and only if $\tilde R^{ij} \neq 0$. Note, that even if $R=0$ we do not obtain in general that the traceless part of the Ricci curvature also vanishes. As a consistency check we can count the number of degrees of freedom. The trace component of the metric EOM fixes $\lambda$, the Hamiltonian constraint and the momentum constraint fixes $N$ and $N^k$ and finally the constraint $C_R$ fixes the trace of the metric. Using the remaining spatially gauge invariance we can fix 3 further components of the metric so that we are left with two traceless components of the spatial metric, $h_{ij}$.

\subsubsection*{Gravitational waves}
Let us consider the consequences for the gravitational waves around a generic background. 
\begin{align}
    h_{ij} = \bar h_{ij} + \delta h_{ij}
\end{align}
We use our gauge symmetry to fix the Donder-gauge, $D_k \delta h^{k}_j=0$. From the constraint EOM we obtain
\begin{align}
    \bar D^2 \delta h = - \tilde R^{ij} \delta \tilde h_{ij}
\end{align}
where we have used the constraint EOM at the background level. For simplicity let us focus first only on the terms with the highes number of derivatives acting on $\delta h_{ij}$.

So far we have not used a specific model for the Lagrangian $\mathcal{L}$. Let us now consider the K-essence model. {\color{red} Check the first EOM  }
\begin{align}
     \frac{1}{\sqrt{h}}\frac{\delta (\sqrt{h} N \mathcal{L})}{\delta N} =&  - \frac{1}{2} ( K_{ij} K^{ij} - K^2) + P - 2  P_X \frac{1}{N^2} = 0, \\
    \frac{\delta \mathcal{L}}{\delta N^k} =& D_k K - D_j K_k^j =0, \\
    \frac{1}{\sqrt{h}} \frac{\delta (\sqrt{h} N \mathcal{L})}{\delta h_{ij}} =& \frac{1}{2} N h^{ij} \mathcal{L} - \frac{1}{2} N R^{ij}  - \frac{1}{2 \sqrt{h}} \frac{\md }{\md t} \left(  \sqrt{h} K^{ij} \right) - \frac{1}{2}   ( K^{i k} D^j N_k + K^{j k} D^j N_k ) \nonumber \\
    & + \frac{1}{2} D_k (  K^{ij} N^k) -  \bar N K^{ki} K_k^j + N K K^{ij} + \frac{1}{2\sqrt{h}} \frac{\md}{\md t}  \left( \sqrt{h}  K h^{ij}\right) \nonumber \\
    & +  h^{ij} K D_k N^k - \frac{1}{2} h^{ij} D_k ( K N^k) + \frac{1}{2} D^i D^j N -  \frac{1}{2} h^{ij} D^2 N  
\end{align}
Considering only the highest number of derivatives (spatial and time) we obtain for the perturbation of the lapse function and the shift vector
\begin{align}
    &\frac{\delta N}{\bar N} \left( \bar K_{ij} \bar K^{ij} - \bar K^2 + P_X \frac{2}{\bar N^2} + 4 P_{XX} \frac{1}{\bar N^4} \right) - \bar K^i_j \frac{1}{2 \bar N} \left( \delta \dot h^{j}_i - \bar D_i \delta N^j  -\bar  D^j \delta N_i \right) - \frac{\bar K }{\bar N}\bar D_k \delta N^k = 0, \\
    & - \bar D_k \bar D_m \delta N^m - \frac{1}{2} \left( \bar D_j \delta \dot h^j_k - \bar D^2 \delta N_k - \bar D_j \bar D_k \delta N^j  \right) =0
\end{align}
So that we can see that $\delta N$ is of first order in derivative of $\delta h_{ij}$ while $\bar D_j \delta N^k$ is of zero order in derivative.
Using this property 
\begin{align}
    \delta N =  \left(  P -  P_X \frac{1}{N^2} + 2 P_{XX} \frac{1}{\bar N^4} \right)^{-1} \bar K^i_j \delta \dot h^j_i, \qquad \delta N^k=0 
\end{align}
Expanding up to leading order in derivatives we obtain for the variation wrt $h_{ij}$
\begin{align}
    \frac{1}{\sqrt{h}}\frac{\delta (\sqrt{h} N \mathcal{L})}{\delta h_{ij}} =& - \frac{1}{2} \bar N \delta R^{ij} - \frac{1}{2}\delta \dot K^{ij} + \frac{1}{2} \bar N^k \bar D_k \delta K^{ij} + \frac{1}{2} \bar h^{ij} \delta \dot K - \frac{1}{2} \bar h^{ij} \bar N^k \bar D_k \delta K \nonumber \\
    & + \frac{1}{2} \bar D^i \bar D^j \delta N - \frac{1}{2} \bar h^{ij} \bar D^2 \delta N 
\end{align}
Therefore, the solution for $\lambda$ yields
\begin{align}
    \bar D^2 \delta \lambda = \frac{1}{2} \delta \dot K  - \frac{1}{2} \bar N^k \bar D_k \delta K - \frac{1}{2} \bar D^2 \delta N
\end{align}
For the traceless EOM we obtain at leading order 
\begin{align}
    &  - \frac{1}{2} \bar N \delta R^{ij} - \frac{1}{2}\delta \dot K^{ij} + \frac{1}{2} \bar N^k \bar D_k \delta K^{ij} + \frac{1}{2} \bar h^{ij} \delta \dot K - \frac{1}{2} \bar h^{ij} \bar N^k \bar D_k \delta K + \frac{1}{2} \bar D^i \bar D^j \delta N - \frac{1}{2} \bar h^{ij} \bar D^2 \delta N \nonumber \\
    & - \frac{1}{2} \bar h^{ij} \big(  \delta \dot K -  \bar N^k \bar D_k \delta K - \bar D^2 \delta N \big) + \bar D^i \bar D^j \delta \lambda -  \bar \lambda \delta  R^{ij} \nonumber \\
    =&   - \frac{1}{2}\delta \dot K^{ij} + \frac{1}{2} \bar N^k \bar D_k \delta K^{ij} - ( \frac{1}{2} \bar N +\bar \lambda ) \delta R^{ij}  + \bar D^i \bar D^j \delta \lambda + \frac{1}{2} \bar D^i \bar D^j \delta N \nonumber \\
    =& - \frac{1}{2}\delta \dot K^{ij} + \frac{1}{2} \bar N^k \bar D_k \delta K^{ij} - ( \frac{1}{2} \bar N +\bar \lambda ) \delta R^{ij}  + \frac{1}{2} \frac{\bar D^i \bar D^j}{\bar D^2} \big( \delta \dot K - \bar N^k \bar D_k \delta K \big)   =0
\end{align}
We can see that the constraint alters the sound speed of the gravitational waves. In particular, the sound speed depends directly on the Lagrange parameter. 

If the background has non-vanishing traceless components for the extrinsic curvature or the metric then the expression will become quite cumbersome and we have to be more careful how to treat it since $\delta N \propto \bar K^{ij} \delta \dot h_{ij}$. Further, the EOM will include non-local terms for the gravitational wave propagation. Note, that due to the broken full diffeomorphism invariance even in spherical symmetric configurations the shift vector cannot be always removed by performing a coordinate transformation. On the other hand, if $\bar K_{ij} \propto \delta_{ij}$ we recover the standard EOM for gravitational waves on a generic background up to a modified propagation speed. 

\subsection{Comparison of linear perturbation theory}

As mentioned before, at the FLRW background the constraint vanishes trivially and, therefore, we end up with a dynamical degree of freedom. In the following, let us test if this peculiar property is consistent or signalizes a breakdown of perturbation theory  by comparing the background evolution to a generic spherically time dependent background where we take the FLRW limit at the end. In order to have a scalar degree of freedom at the linear level we will add an additional minimally coupled scalar field for simplicity
\begin{align}
    S_m = \int \md^3 \md t\, \sqrt{h} N \Big( \frac{1}{2 N^2} (\dot \chi - N^k \partial_k \chi)^2 - \frac{1}{2} h^{ij} \partial_i \chi \partial_j \chi -  V(\chi) \Big)
\end{align}

\subsubsection*{FLRW background}
At the FLRW background the EOM are not impacted by the constraint and we recover the usual EOM of K-essence plus the minimally coupled scalar field
\begin{align}
    3 H^2 =& - P + 2 P_X \frac{1}{N^2} + \frac{1}{2 N^2} \dot \chi^2 +  V_\chi, \\
    3 H^2 + 2 \dot H =& ..., \\
    \ddot \chi + 3 H \dot \chi + V_\chi =& 0.
\end{align}

\subsubsection*{Spherically symmetric background}
The most general ansatz for a spherical symmetric background is given by
\begin{align}
    \md s^2 = - N^2(t,r) \md t^2 + ( B(t,r) \md t + a(t) F(t,r) \md r)^2 + a(t)^2 r^2 \md \Omega^2
\end{align}
Note, that we cannot remove $B(t,r)$ via a coordinate transformation since the transformation is not inside the restricted spatial diffeomorphism class. 

Let us first consider the constraint ({\color{red} check it })
\begin{align}
    R = \frac{2}{r^2} - \frac{2}{r^2 F^2} + \frac{4 F^\prime }{r F^3} =0
\end{align}
which leads to the solution
\begin{align}
    F(t,r)  = \pm \frac{1}{\sqrt{1+\frac{r_0(t)}{r}}}
\end{align}
where $r_0(t)$ is an integration constant. In the following we will consider the positive solution. For $r_0(t)\neq 0$ we can study the case of black hole solutions. In the following we will set the integration constant to zero to study the cosmological limit. 

Note, that due to the spherical symmetry the EOM for the metric, lapse and shift vector are the same as in the original K-essence theory except for the $r r$ component of the metric equations which can be used to fix the Lagrange multiplier. However, in the absence of the constraint the $r r$ component is anyway equivalent to the $\theta\theta$ or $\phi \phi$ component and, therefore, we obtain essentially the same background EOM plus the constraint EOM. 

From the momentum constraint
\begin{align}
     \frac{1}{2} \left( D_j K - D_k K^k_j \right) - \frac{1}{N} \left(\dot \chi - N^k \partial_k \chi \right)\partial_j \chi = 0, 
\end{align}

\fi

\if{}
\section{Minimally modified gravity linear in lapse}
Let us consider the case where the model is linear in the lapse
\begin{align}
    \mathcal{H} = N \mathcal{H}_0( \pi^{ij},D_k, h_{ij},t) + V(\pi^{ij}, D_k, h_{ij},t)
\end{align}
The constraint itself can be expressed as
\begin{align}
    C = C(\pi^{ij}, h_{ij},D_k,t)
\end{align}
In the following let us assume for simplicity that the constraint and the Hamiltonian does not depend on time explicitly.

\subsection{Breakdown of perturbation theory}
In order to compare the case to the previous analysis let us assume that the constraint is trivially fulfilled on the FLRW background. As one example, we can consider the same constraint $C_R= \sqrt{h} R$. However, contrary to the previous case the FLRW background is still not dynamical. This is coming from the fact that at the background the Hamiltonian constraint $\mathcal{H}_0$ will commute with itself and, therefore, the both constraint $\pi_N$ and $\mathcal{H}_0$ form again a pair of two first class constraints removing the scalar degree of freedom at the background level. 

On the other hand, if the constraint also vanishes at linear level as $C= R_{ij} R^{ij}$ then the constraint is fulfilled trivially even at the linear level and the perturbation theory breaks down since we obtain a scalar degree of freedom at linear level. To see this explicitly, we can consider the simple case of 
\begin{align}
    \mathcal{H} = N \mathcal{H}_{GR} + c_1 R + c_2 \pi 
\end{align}
and for the constraint 
\begin{align}
    C_R = R_{ij} R^{ik} R^j_k
\end{align}
{\color{red} The example may not be important or could be shifted into the appendix}

\subsection{FLRW background}

\subsection{Impact of the constraints}
To get an impression of the impact of the different constraint let us consider two simple examples for the constraint
\begin{align}
    C_R = \sqrt{h} R, \qquad
    C_K = D_k D^k \pi
\end{align}
To simplify the discussion we consider for the Hamiltonian 
\begin{align}
    \mathcal{H} =& \frac{N}{\sqrt{h}} \left( \sqrt{h} c_0 \pi + c_1 \pi^{ij} \pi_{ij} + c_2 \pi^2 \right)     + \frac{1}{\sqrt{h}} \left( \sqrt{h} c_4 \pi + c_5 \pi_{ij} \pi^{ij} + c_6 \pi^2 \right) \nonumber \\
    &+ \sqrt{h} N f(h_{ij},D_k) + \sqrt{h} g(h_{ij},D_k)
\end{align}
Since the constraints vanish on the FLRW background identically they do not impact the EOM at the background level. So for both constraints we obtain 
\begin{align}
 & (c_1 + 3 c_2 + c_5 + 3 c_6 )^2 \rho_m + 3 (c_1 + 3 c_2) H^2 + 3 (  c_0 (c_5 + 3 c_6) - c_4 (c_1 + 3c_2) ) H  \nonumber \\
 & + (c_1 + 3 c_2 + c_5 + 3 c_6 ) f_\Lambda - \frac{3}{4} (c_0 + c_4) ( - c_4( c_1 + 3 c_2) + c_0 (c_1 + 3 c_2 + 2c_5 + 6 c_6 ) )=0, \\
 & 2 \dot H + 3 H^2 = \frac{1}{4} \left( 3 c_0^2 + 6 c_0 c_4 + 3 c_4^2 - 4( c_1 +3 c_2 +  c_5 + 3 c_6 ) (f_\Lambda + g_\Lambda)  \right), \\
 & \dot \rho_m + 3 H \rho_m =0
\end{align}

\subsubsection*{Curvature constraint}
Let us start with the curvature constraint and compare it with the other approach. 
In this case, performing the Legendre transformation the Lagrangian is given by
\begin{align}
    \mathcal{L}_{tot}=& \sqrt{h} N \mathcal{L} + \lambda C_R  \nonumber \\
    =&  \sqrt{h} N \left( \frac{1}{\tilde c_1} \mathcal{K}_{ij} \mathcal{K}^{ij} - \frac{\tilde c_2 }{\tilde c_1 (\tilde c_1 + 3 \tilde c_2)} \mathcal{K}^2  - f(h_{ij},D_k) \right) - \sqrt{h} g(h_{ij},D_k) + \lambda C_R
\end{align}
where
\begin{align}
    \tilde c_1 = c_1 + \frac{c_5}{N}, \qquad \tilde c_2 = c_2 + \frac{c_6}{N}, \qquad \mathcal{K}_{ij} = K_{ij} - \frac{1}{2N} \left( c_0 N + c_4 \right) h_{ij} 
\end{align}
As before, the Lagrange multiplier can be obtained by taking the trace of the spatial metric equation yielding
\begin{align}
    D^2 \lambda = \frac{h_{ij}}{2 \sqrt{h}} \frac{\delta (\sqrt{h} N \mathcal{L})}{\delta h_{ij}}
\end{align}
Therefore, the Lagrange multiplier can be obtained by solving the elliptic equation. 

At the linear level around the FLRW background we obtain after integrating out the non-dynamical perturbations $\psi$, $\delta N$, $\delta j$, $\delta j_0$ and $\delta \lambda$ (see appendix)
\begin{align}
    \delta S = \int \md^3k\md t\, a^3 \frac{z^2}{2} \left( \delta \dot \sigma^2 - \left( c_s^2 \frac{k^2}{a^2} + M^2 \right) \delta \sigma^2 \right)
\end{align}
where
\begin{align}
    z^2 =& \frac{(-c_4(c_1 + 3c_2) + c_0 (c_5 + 3 c_6) + 2 (c_1 + 3 c_2) H)^2}{(c_1 + 2 c_2 + c_5 + 2c_6) (c_1 + 3c_2 + c_5 + 3 c_6)^2}, \\
    c_s^2 =&  \frac{(c_1 + 2 c_2 + c_5 + 2c_6) (c_1 + 3c_2 + c_5 + 3 c_6)^2}{(-c_4(c_1 + 3c_2) + c_0 (c_5 + 3 c_6) + 2 (c_1 + 3 c_2) H)^2} \rho_m, \\
    M^2 =& \frac{(c_1+c_5) (c_1+3c_2+c_5+3c_6) \rho_m}{(-c_4(c_1 + 3c_2) + c_0 (c_5 + 3 c_6) + 2 (c_1 + 3 c_2) H)^2} \nonumber \\
    & \times \left(4 (c_1+3c_2+c_5+3c_6)^2 \rho_m - 8 (c_1+3c_2) \dot H   \right)
\end{align}
Therefore, in general the dust will start to propagate. There is one specific case, namely when $c_1 + 2c_2 + c_5 + 2 c_6=0$ leading to $c_s^2=0$ and $z^2 \rightarrow \infty$. Redoing the analysis we obtain that there is actually only a half degree of freedom at all (see appendix for details). Note, that the case of the GR is inside this class. 

\subsubsection*{Flat slicing constraint}
In the case of the constraint $C_K$ the Lagrangian is given by
\begin{align}
    \sqrt{h} N \mathcal{L} = \sqrt{h} N \left( \frac{1}{\tilde c_1} \mathcal{K}_{K,ij} \mathcal{K}^{ ij}_K - \frac{\tilde c_2 }{\tilde c_1 (\tilde c_1 + 3 \tilde c_2)} \mathcal{K}^{ 2}_K  - f(h_{ij},D_k) \right) - \sqrt{h} g(h_{ij},D_k)
\end{align}
where
\begin{align}
    \mathcal{K}_{K,ij}  = K_{ij} - \frac{1}{2N} \left( c_0 N + c_4 \right) h_{ij} + \frac{1}{2N} h_{ij} D^2 \lambda
\end{align}
The Lagrange parameter can be obtained by solving the constraint EOM itself yielding
\begin{align}
    D^2 \left( \frac{1}{\tilde c_1 + 3 \tilde c_2 } \mathcal{K} \right) =0
\end{align}
There is one interesting specific subcase $c_4=0$, and $3 c_5= -c_6$ in which
\begin{align}
    D^2 \left( K + \frac{3}{2N} D^2 \lambda \right)=0
\end{align}
Therefore, the Lagrange multiplier is again solved up to an integration constant. 
\fi

\section{Conclusion}
\label{sec:Conclusion}
In this paper we have analyzed minimally modified gravity models with a dynamical FLRW background evolution. This can be obtained by imposing auxiliary constraints which vanish identically at the FLRW background (see also \cite{DeFelice:2015hla,DeFelice:2015moy,Ganz:2021hmp}) so that we obtain the background evolution from the original model.
This imposes conditions on the form of the constraints. In particular, we need to ensure that the constraints are not trivial identities at the linear order since it leads to a breakdown of perturbation theory. 

While it is a priori also possible to construct this type of models without the need of auxiliary constraints this leads to highly non-standard cosmologies requiring for instance a trivial Hamiltonian constraint at the background level. These models can also suffer under a breakdown of linear perturbation theory around FLRW. 
\bigskip

In the next part we studied the phenomenological consequences. 
For generic backgrounds the auxiliary constraints can lead to a non-standard dispersion relation for the gravitational waves and is, in general, highly sensitive to the form of the constraints.

As a next step we focused on the linear perturbation around the FLRW background including dust to have a dynamical scalar degree of freedom at the linear level. For two classes of constraints (one primary constraint, which does not depend on the momenta of the metric, or two generic primary constraints) the perturbations around FLRW are not very sensitive to the specific details of the constraints. In the first case depending on the original model the constraint will lead to a modification of the effective gravitational constant for the dust component and for models outside the GLPV class it can lead to a non-vanishing sound speed. This is similar to the second case where the two primary constraints will, in general, except for some specific cases always provide a non-vanishing sound speed for dust which is highly constrained by observations. 

For the tensor modes the details of the constraints can become important. In this case the tensor sector will depend on the background value of the Lagrange multiplier which is, however, not fixed since the constraint is trivially fulfilled at the background level. However, we can avoid this ambiguity by further restricting the form of the constraints to $C= \sqrt{h} D_k C^k$ so that the EOM are invariant under a time dependent shift of the Lagrange multiplier. In that case the tensor sector will not be impacted by the constraint up to linear order but instead we recover the same result as for the original model prior to the constraint.
\bigskip

Summarizing, constructing MMG models by imposing auxiliary constraints provides a new playground leading to interesting new phenomenological features. In future it might be interesting to study the consequences of these type of models in the case of black holes or other backgrounds where the constraint do not vanish trivially.

\acknowledgments
It is a pleasure to thank Chunshan Lin for useful discussions. A.G. is supported  by the grant No. UMO-2021/40/C/ST9/00015 from the National Science Centre, Poland.

\appendix

\section{Non-trivial background}
\label{app:Non_trivial_background}
In order to get a better understanding of the dynamical degree of freedom at the FLRW background let us consider a generic stationary spherical symmetric background in which the constraint is not a trivial identity
\begin{align}
    \md s^2 = - N(t,r)^2 \md t^2 + F(t,r)^2 a(t)^2 \md r^2 + a(t)^2 r^2 \md \Omega^2.
\end{align}
In order to simplify the discussion let us consider the constraint $C=\sqrt{h}R$ which leads to
\begin{align}
    - F(t,r) + F(t,r)^3 + 2 r \partial_r F(t,r) =0.
\end{align}
Solving it we obtain
\begin{align}
    F(t,r) = \frac{1}{\sqrt{\frac{\kappa(t)}{r} + 1}}.
\end{align}
By imposing proper boundary conditions at spatial infinity we can set $\kappa(t)=0$ recovering the standard result. 

In that case the equations of motion for the toy model in \eqref{eq:Toy_model} and \eqref{eq:Toy_model_constraint} simplify to
\begin{align}
    &3 \frac{\dot a(t)^2}{a(t)^2} + N(t,r)^2 P - 2 P^\prime=0, \\
   & a(t)^2 N(t,r)^3 P + N(t,r) ( \dot a(t)^2 + 2 a(t) \ddot a(t) ) \nonumber \\
   &- \frac{2}{r} N(t,r)^2 ( \partial_r N(t,r) + 2 \partial_r \lambda(t,r) ) -  2 \dot a(t) a(t)\dot N(t,r)=0, \\
   & a(t)^2 N(t,r)^2 P + \dot a(t)^2 + 2 a(t) \ddot a(t) - 2 a(t) \dot a(t) \frac{\dot N(t,r)}{N(t,r)} =0.
\end{align}
From the Hamiltonian constraint we obtain that if we do not want to constrain the form of the free function $P(1/N(t,r)^2)$ we need to constrain $N(t,r)=N(t)$. Using it we recover the usual equation of motion from the flat FLRW background, i.e.
\begin{align}
    3 H^2 +  P - \frac{2}{N^2} P^\prime =0, \\
    3 H^2 + 2 \dot H + P =0, \\
    \partial_r \lambda = 0,
\end{align}
where $H(t)= \dot a(t)/(N(t) a(t))$. Therefore, we can note that it is possible to obtain the flat FLRW solutions in a smooth limit by imposing proper boundary conditions for $F(t,r)$ at spatial infinity. 

\section{Dynamical FLRW background without auxiliary constraints}
\label{app:Dynamical_FLRW}
The dynamical FLRW background can also be achieved in models without the presence of auxiliary constraints. However, in general, the degeneracy constraints are quite cumbersome to solve. In models like Cuscuton \cite{Afshordi:2006ad,Gomes:2017tzd} etc. the constraint structure is fundamental different. Besides the usual six first class constraints related to the spatial diffeomorphism invariance there is an additional first class constraint related to $\pi_N$ and two second class constraints $\mathcal{H}_0$ and a new tertiary one $C$. For SCG models as
\begin{align}
\label{eq:SCG_ansatz}
    S = \int \md^4x\,  \sqrt{h}N L(N,K_{ij},h_{ij},D_k,t)
\end{align}
degenerate conditions have been derived in \cite{Gao:2019twq} in order to obtain a MMG theory.
Note, that while, in general, $\pi_N$ might not be anymore first class there will be a specific combination $\hat \pi_N$ which remains first class besides the six first class constraints coming from the spatial diffeomorphism invariance \cite{Gao:2019twq}. The total Hamiltonian at the FLRW background can be written as
\begin{align}
    H_T\vert_{FLRW} = \int \md^3x\, \left( \mathcal{H}(a,p_a,N) + u_N \hat \pi_N + u_0 \mathcal{H}_0 + u_C C \right)\vert_{FLRW}
\end{align}
where we have used that the momentum constraint is trivial at the FLRW background. 
Therefore, even if the tertiary constraints vanish at the background level $C\vert_{FLRW}=0$ this will not lead to a dynamical background since we are left with the two constraints $\hat \pi_N$ and the Hamiltonian constraint $\mathcal{H}_0$. Since at the FLRW background all constraints will commute with itself  $\mathcal{H}_0$ becomes first class since $\hat \pi_N$ is first class. Therefore, it is quite challenging to obtain a dynamical FLRW background.
Besides requiring a trivial tertiary constraint $C\vert_{FLRW}=0$ we have to require that either $\hat \pi_N$ or $\mathcal{H}_0$ become trivial at the background level. 
The most straightforward way is to construct a Hamiltonian constraint which becomes trivially at the FLRW background leading to a non-standard cosmology. 

As discussed in \cite{Mukohyama:2019unx} models of the form
\begin{align}
    H = \int \md^3x\,  \left[ \mathcal{V}(h_{ij},\pi^{ij},D_k) + N \mathcal{H}_0(h_{ij},\pi^{ij},D_k) - 2 \sqrt{h} N^k D^j \left( \frac{\pi_{kj}}{\sqrt{h}} \right) \right]
\end{align}
where
\begin{align}
    \{ \mathcal{H}_0(x), \mathcal{H}_0(y) \} \approx 0
\end{align}
do just have two tensor degrees of freedom. Therefore, if $\mathcal{H}_0\vert_{FLRW} =0$ we obtain a minimally modified gravity model with a dynamical FLRW background.
One easy example is given by the following toy model
\begin{align}
    \mathcal{L} = \frac{1}{2} \sqrt{h} N^2 (K^{ij} K_{ij} - K^2) - N \mathcal{H}_0(h_{ij},D_k) 
\end{align}
where $\mathcal{H}_0(h_{ij},D_k)$ can be any generic function as long as it does vanish at the FLRW background as $\mathcal{H}_0= \sqrt{h} R$. However, if we add matter the Hamiltonian constraint is, in general, not anymore trivial. Therefore, including conventional matter the theory has still just one dynamical degree of freedom at the background level. Furthermore, the Hamiltonian constraint enforces that at the background level the matter energy density vanishes. All in all, these models are highly pathological. 

Note, that if we consider a Hamiltonian constraint which is also trivial at the linear level as $\mathcal{H}_0 = \sqrt{h} R^{ij} R_{ij}$ then the linear perturbations around the FLRW break down, i.e. there are three degress of freedom (2 tensor + 1 scalar) at the linear level. It is the same issue which we have discussed in the case of the auxiliary constraints in section \ref{subsec:Breakdown}. 
\bigskip

Last, it might be interesting to check if it is possible to obtain MMG models with dynamical dark energy by generalizing the ansatz \eqref{eq:SCG_ansatz} by for instance including $\dot N$ as in \cite{Lin:2020nro} or breaking the spatial diffeomorphism invariance to check if there are viable models in the context of cosmology.

\section{Schultz-Sorkin action}
\label{sec:Schultz-Sorkin}
In order to describe dust we use the Schultz-Sorkin action \cite{schutz1977variational}. Using the implementation as in \cite{Felice_2016}

\begin{align}
    S_{\mathrm{mat}} =& - \int \md^4x\, \left( \sqrt{-g} \rho_m(n) + J^\alpha \partial_\alpha \varphi \right), \\
    \rho_m(n) =& \mu_0 n, \\
    n=& \sqrt{\frac{J^\alpha J^\beta g_{\alpha\beta}}{g}},
\end{align}
where $J^\alpha$ is a vector of weight one and  $\varphi$, $n$ and $\rho_m$ are scalar fields. Up to linear order the vector $J^\alpha$ and the scalar field $\varphi$ can be expressed as
\begin{align}
    J^0 =& \mathcal{N}_0 + \delta j_0, \\
    J^k = & \delta^{kj} \partial_j \delta j, \\
    \varphi = & - \mu \int^t \md \tau N(\tau) - \mu_0 v_m,
\end{align}
where $\mathcal{N}_0$ is the number of dust particles at the background level with $a^3 \rho_m = \mu_0 \mathcal{N}_0 $.
Further, it is convenient to replace $\delta j_0$ with the gauge invariant matter overdensity $\delta_m$ via
\begin{align}
     \frac{\delta j_0}{\mathcal{N}_0} = \delta_m - 3 H v_m + 3 \xi .
\end{align}

\bibliography{bibliography}

\providecommand{\href}[2]{#2}\begingroup\raggedright\begin{thebibliography}{10}

\bibitem{Aoki:2018zcv}
K.~Aoki, C.~Lin and S.~Mukohyama, \emph{{Novel matter coupling in general
  relativity via canonical transformation}},
  \href{https://doi.org/10.1103/PhysRevD.98.044022}{\emph{Phys. Rev. D}
  {\bfseries 98} (2018) 044022},
  [\href{https://arxiv.org/abs/1804.03902}{{\ttfamily 1804.03902}}].

\bibitem{Lin:2018mip}
C.~Lin, \emph{{The Self-consistent Matter Coupling of a Class of Minimally
  Modified Gravity Theories}},
  \href{https://doi.org/10.1088/1475-7516/2019/05/037}{\emph{JCAP} {\bfseries
  05} (2019) 037}, [\href{https://arxiv.org/abs/1811.02467}{{\ttfamily
  1811.02467}}].

\bibitem{Iyonaga:2018vnu}
A.~Iyonaga, K.~Takahashi and T.~Kobayashi, \emph{{Extended Cuscuton:
  Formulation}},
  \href{https://doi.org/10.1088/1475-7516/2018/12/002}{\emph{JCAP} {\bfseries
  12} (2018) 002}, [\href{https://arxiv.org/abs/1809.10935}{{\ttfamily
  1809.10935}}].

\bibitem{Gao:2019twq}
X.~Gao and Z.-B. Yao, \emph{{Spatially covariant gravity theories with two
  tensorial degrees of freedom: the formalism}},
  \href{https://doi.org/10.1103/PhysRevD.101.064018}{\emph{Phys. Rev. D}
  {\bfseries 101} (2020) 064018},
  [\href{https://arxiv.org/abs/1910.13995}{{\ttfamily 1910.13995}}].

\bibitem{Lin:2017oow}
C.~Lin and S.~Mukohyama, \emph{{A Class of Minimally Modified Gravity
  Theories}}, \href{https://doi.org/10.1088/1475-7516/2017/10/033}{\emph{JCAP}
  {\bfseries 10} (2017) 033},
  [\href{https://arxiv.org/abs/1708.03757}{{\ttfamily 1708.03757}}].

\bibitem{Mukohyama:2019unx}
S.~Mukohyama and K.~Noui, \emph{{Minimally Modified Gravity: a Hamiltonian
  Construction}},
  \href{https://doi.org/10.1088/1475-7516/2019/07/049}{\emph{JCAP} {\bfseries
  07} (2019) 049}, [\href{https://arxiv.org/abs/1905.02000}{{\ttfamily
  1905.02000}}].

\bibitem{DeFelice:2020eju}
A.~De~Felice, A.~Doll and S.~Mukohyama, \emph{{A theory of type-II minimally
  modified gravity}},
  \href{https://doi.org/10.1088/1475-7516/2020/09/034}{\emph{JCAP} {\bfseries
  09} (2020) 034}, [\href{https://arxiv.org/abs/2004.12549}{{\ttfamily
  2004.12549}}].

\bibitem{Yao:2020tur}
Z.-B. Yao, M.~Oliosi, X.~Gao and S.~Mukohyama, \emph{{Minimally modified
  gravity with an auxiliary constraint: A Hamiltonian construction}},
  \href{https://doi.org/10.1103/PhysRevD.103.024032}{\emph{Phys. Rev. D}
  {\bfseries 103} (2021) 024032},
  [\href{https://arxiv.org/abs/2011.00805}{{\ttfamily 2011.00805}}].

\bibitem{Tasinato:2020fni}
G.~Tasinato, \emph{{Symmetries for scalarless scalar theories}},
  \href{https://doi.org/10.1103/PhysRevD.102.084009}{\emph{Phys. Rev. D}
  {\bfseries 102} (2020) 084009},
  [\href{https://arxiv.org/abs/2009.02157}{{\ttfamily 2009.02157}}].

\bibitem{Aoki:2021zuy}
K.~Aoki, F.~Di~Filippo and S.~Mukohyama, \emph{{Non-uniqueness of massless
  transverse-traceless graviton}},
  \href{https://doi.org/10.1088/1475-7516/2021/05/071}{\emph{JCAP} {\bfseries
  05} (2021) 071}, [\href{https://arxiv.org/abs/2103.15044}{{\ttfamily
  2103.15044}}].

\bibitem{Lin:2020nro}
J.~Lin, Y.~Gong, Y.~Lu and F.~Zhang, \emph{{Spatially covariant gravity with a
  dynamic lapse function}},
  \href{https://doi.org/10.1103/PhysRevD.103.064020}{\emph{Phys. Rev. D}
  {\bfseries 103} (2021) 064020},
  [\href{https://arxiv.org/abs/2011.05739}{{\ttfamily 2011.05739}}].

\bibitem{Ganz:2021hmp}
A.~Ganz and C.~Lin, \emph{{Structure Formation in the Effective Field Theory of
  Holographic Dark Energy}},
  \href{https://arxiv.org/abs/2109.07420}{{\ttfamily 2109.07420}}.

\bibitem{Bartolo:2021wpt}
N.~Bartolo, A.~Ganz and S.~Matarrese, \emph{{Cuscuton Inflation}},
  \href{https://arxiv.org/abs/2111.06794}{{\ttfamily 2111.06794}}.

\bibitem{Frieman:2008sn}
J.~Frieman, M.~Turner and D.~Huterer, \emph{{Dark Energy and the Accelerating
  Universe}},
  \href{https://doi.org/10.1146/annurev.astro.46.060407.145243}{\emph{Ann. Rev.
  Astron. Astrophys.} {\bfseries 46} (2008) 385--432},
  [\href{https://arxiv.org/abs/0803.0982}{{\ttfamily 0803.0982}}].

\bibitem{Clifton:2011jh}
T.~Clifton, P.~G. Ferreira, A.~Padilla and C.~Skordis, \emph{{Modified Gravity
  and Cosmology}},
  \href{https://doi.org/10.1016/j.physrep.2012.01.001}{\emph{Phys. Rept.}
  {\bfseries 513} (2012) 1--189},
  [\href{https://arxiv.org/abs/1106.2476}{{\ttfamily 1106.2476}}].

\bibitem{Joyce:2014kja}
A.~Joyce, B.~Jain, J.~Khoury and M.~Trodden, \emph{{Beyond the Cosmological
  Standard Model}},
  \href{https://doi.org/10.1016/j.physrep.2014.12.002}{\emph{Phys. Rept.}
  {\bfseries 568} (2015) 1--98},
  [\href{https://arxiv.org/abs/1407.0059}{{\ttfamily 1407.0059}}].

\bibitem{Afshordi:2006ad}
N.~Afshordi, D.~J.~H. Chung and G.~Geshnizjani, \emph{{Cuscuton: A Causal Field
  Theory with an Infinite Speed of Sound}},
  \href{https://doi.org/10.1103/PhysRevD.75.083513}{\emph{Phys. Rev. D}
  {\bfseries 75} (2007) 083513},
  [\href{https://arxiv.org/abs/hep-th/0609150}{{\ttfamily hep-th/0609150}}].

\bibitem{Gomes:2017tzd}
H.~Gomes and D.~C. Guariento, \emph{{Hamiltonian analysis of the cuscuton}},
  \href{https://doi.org/10.1103/PhysRevD.95.104049}{\emph{Phys. Rev. D}
  {\bfseries 95} (2017) 104049},
  [\href{https://arxiv.org/abs/1703.08226}{{\ttfamily 1703.08226}}].

\bibitem{Crisostomi:2017ugk}
M.~Crisostomi, K.~Noui, C.~Charmousis and D.~Langlois, \emph{{Beyond Lovelock
  gravity: Higher derivative metric theories}},
  \href{https://doi.org/10.1103/PhysRevD.97.044034}{\emph{Phys. Rev. D}
  {\bfseries 97} (2018) 044034},
  [\href{https://arxiv.org/abs/1710.04531}{{\ttfamily 1710.04531}}].

\bibitem{DeFelice:2018ewo}
A.~De~Felice, D.~Langlois, S.~Mukohyama, K.~Noui and A.~Wang,
  \emph{{Generalized instantaneous modes in higher-order scalar-tensor
  theories}}, \href{https://doi.org/10.1103/PhysRevD.98.084024}{\emph{Phys.
  Rev. D} {\bfseries 98} (2018) 084024},
  [\href{https://arxiv.org/abs/1803.06241}{{\ttfamily 1803.06241}}].

\bibitem{Ganz:2019vre}
A.~Ganz, N.~Bartolo and S.~Matarrese, \emph{{Towards a viable effective field
  theory of mimetic gravity}},
  \href{https://doi.org/10.1088/1475-7516/2019/12/037}{\emph{JCAP} {\bfseries
  12} (2019) 037}, [\href{https://arxiv.org/abs/1907.10301}{{\ttfamily
  1907.10301}}].

\bibitem{Blas:2009yd}
D.~Blas, O.~Pujolas and S.~Sibiryakov, \emph{{On the Extra Mode and
  Inconsistency of Horava Gravity}},
  \href{https://doi.org/10.1088/1126-6708/2009/10/029}{\emph{JHEP} {\bfseries
  10} (2009) 029}, [\href{https://arxiv.org/abs/0906.3046}{{\ttfamily
  0906.3046}}].

\bibitem{doi:10.1063/1.1665613}
D.~Lovelock, \emph{The einstein tensor and its generalizations},
  \href{https://doi.org/10.1063/1.1665613}{\emph{Journal of Mathematical
  Physics} {\bfseries 12} (1971) 498--501},
  [\href{https://arxiv.org/abs/https://doi.org/10.1063/1.1665613}{{\ttfamily
  https://doi.org/10.1063/1.1665613}}].

\bibitem{doi:10.1063/1.1666069}
D.~Lovelock, \emph{The four‐dimensionality of space and the einstein tensor},
  \href{https://doi.org/10.1063/1.1666069}{\emph{Journal of Mathematical
  Physics} {\bfseries 13} (1972) 874--876},
  [\href{https://arxiv.org/abs/https://doi.org/10.1063/1.1666069}{{\ttfamily
  https://doi.org/10.1063/1.1666069}}].

\bibitem{Gao:2014soa}
X.~Gao, \emph{{Unifying framework for scalar-tensor theories of gravity}},
  \href{https://doi.org/10.1103/PhysRevD.90.081501}{\emph{Phys. Rev. D}
  {\bfseries 90} (2014) 081501},
  [\href{https://arxiv.org/abs/1406.0822}{{\ttfamily 1406.0822}}].

\bibitem{Gao:2014fra}
X.~Gao, \emph{{Hamiltonian analysis of spatially covariant gravity}},
  \href{https://doi.org/10.1103/PhysRevD.90.104033}{\emph{Phys. Rev. D}
  {\bfseries 90} (2014) 104033},
  [\href{https://arxiv.org/abs/1409.6708}{{\ttfamily 1409.6708}}].

\bibitem{SupernovaSearchTeam:1998fmf}
{\scshape Supernova Search Team} collaboration, A.~G. Riess et~al.,
  \emph{{Observational evidence from supernovae for an accelerating universe
  and a cosmological constant}},
  \href{https://doi.org/10.1086/300499}{\emph{Astron. J.} {\bfseries 116}
  (1998) 1009--1038}, [\href{https://arxiv.org/abs/astro-ph/9805201}{{\ttfamily
  astro-ph/9805201}}].

\bibitem{SupernovaCosmologyProject:1998vns}
{\scshape Supernova Cosmology Project} collaboration, S.~Perlmutter et~al.,
  \emph{{Measurements of $\Omega$ and $\Lambda$ from 42 high redshift
  supernovae}}, \href{https://doi.org/10.1086/307221}{\emph{Astrophys. J.}
  {\bfseries 517} (1999) 565--586},
  [\href{https://arxiv.org/abs/astro-ph/9812133}{{\ttfamily
  astro-ph/9812133}}].

\bibitem{DeFelice:2022uxv}
A.~De~Felice, K.-i. Maeda, S.~Mukohyama and M.~C. Pookkillath, \emph{{VCDM and
  Cuscuton}},  \href{https://arxiv.org/abs/2204.08294}{{\ttfamily 2204.08294}}.

\bibitem{Afshordi:2007yx}
N.~Afshordi, D.~J.~H. Chung, M.~Doran and G.~Geshnizjani, \emph{{Cuscuton
  Cosmology: Dark Energy meets Modified Gravity}},
  \href{https://doi.org/10.1103/PhysRevD.75.123509}{\emph{Phys. Rev. D}
  {\bfseries 75} (2007) 123509},
  [\href{https://arxiv.org/abs/astro-ph/0702002}{{\ttfamily
  astro-ph/0702002}}].

\bibitem{DeFelice:2015hla}
A.~De~Felice and S.~Mukohyama, \emph{{Minimal theory of massive gravity}},
  \href{https://doi.org/10.1016/j.physletb.2015.11.050}{\emph{Phys. Lett. B}
  {\bfseries 752} (2016) 302--305},
  [\href{https://arxiv.org/abs/1506.01594}{{\ttfamily 1506.01594}}].

\bibitem{DeFelice:2015moy}
A.~De~Felice and S.~Mukohyama, \emph{{Phenomenology in minimal theory of
  massive gravity}},
  \href{https://doi.org/10.1088/1475-7516/2016/04/028}{\emph{JCAP} {\bfseries
  04} (2016) 028}, [\href{https://arxiv.org/abs/1512.04008}{{\ttfamily
  1512.04008}}].

\bibitem{Vikman:2004dc}
A.~Vikman, \emph{{Can dark energy evolve to the phantom?}},
  \href{https://doi.org/10.1103/PhysRevD.71.023515}{\emph{Phys. Rev. D}
  {\bfseries 71} (2005) 023515},
  [\href{https://arxiv.org/abs/astro-ph/0407107}{{\ttfamily
  astro-ph/0407107}}].

\bibitem{Boruah:2018pvq}
S.~S. Boruah, H.~J. Kim, M.~Rouben and G.~Geshnizjani, \emph{{Cuscuton
  bounce}}, \href{https://doi.org/10.1088/1475-7516/2018/08/031}{\emph{JCAP}
  {\bfseries 08} (2018) 031},
  [\href{https://arxiv.org/abs/1802.06818}{{\ttfamily 1802.06818}}].

\bibitem{Kim:2020iwq}
J.~L. Kim and G.~Geshnizjani, \emph{{Spectrum of Cuscuton Bounce}},
  \href{https://doi.org/10.1088/1475-7516/2021/03/104}{\emph{JCAP} {\bfseries
  03} (2021) 104}, [\href{https://arxiv.org/abs/2010.06645}{{\ttfamily
  2010.06645}}].

\bibitem{PhysRev.166.1263}
R.~A. Isaacson, \emph{Gravitational radiation in the limit of high frequency.
  i. the linear approximation and geometrical optics},
  \href{https://doi.org/10.1103/PhysRev.166.1263}{\emph{Phys. Rev.} {\bfseries
  166} (Feb, 1968) 1263--1271}.

\bibitem{PhysRev.166.1272}
R.~A. Isaacson, \emph{Gravitational radiation in the limit of high frequency.
  ii. nonlinear terms and the effective stress tensor},
  \href{https://doi.org/10.1103/PhysRev.166.1272}{\emph{Phys. Rev.} {\bfseries
  166} (Feb, 1968) 1272--1280}.

\bibitem{Gleyzes:2013ooa}
J.~Gleyzes, D.~Langlois, F.~Piazza and F.~Vernizzi, \emph{{Essential Building
  Blocks of Dark Energy}},
  \href{https://doi.org/10.1088/1475-7516/2013/08/025}{\emph{JCAP} {\bfseries
  08} (2013) 025}, [\href{https://arxiv.org/abs/1304.4840}{{\ttfamily
  1304.4840}}].

\bibitem{Gleyzes:2014dya}
J.~Gleyzes, D.~Langlois, F.~Piazza and F.~Vernizzi, \emph{{Healthy theories
  beyond Horndeski}},
  \href{https://doi.org/10.1103/PhysRevLett.114.211101}{\emph{Phys. Rev. Lett.}
  {\bfseries 114} (2015) 211101},
  [\href{https://arxiv.org/abs/1404.6495}{{\ttfamily 1404.6495}}].

\bibitem{Aoki:2020lig}
K.~Aoki, M.~A. Gorji and S.~Mukohyama, \emph{{A consistent theory of $D \to 4$
  Einstein-Gauss-Bonnet gravity}},
  \href{https://doi.org/10.1016/j.physletb.2020.135843}{\emph{Phys. Lett. B}
  {\bfseries 810} (2020) 135843},
  [\href{https://arxiv.org/abs/2005.03859}{{\ttfamily 2005.03859}}].

\bibitem{Aoki:2020ila}
K.~Aoki, M.~A. Gorji, S.~Mizuno and S.~Mukohyama, \emph{{Inflationary
  gravitational waves in consistent $D \to 4$ Einstein-Gauss-Bonnet gravity}},
  \href{https://doi.org/10.1088/1475-7516/2021/01/054}{\emph{JCAP} {\bfseries
  01} (2021) 054}, [\href{https://arxiv.org/abs/2010.03973}{{\ttfamily
  2010.03973}}].

\bibitem{PhysRevLett.119.161101}
{\scshape LIGO Scientific Collaboration and Virgo Collaboration} collaboration,
  B.~P. Abbott, R.~Abbott, T.~D. Abbott, F.~Acernese, K.~Ackley, C.~Adams
  et~al., \emph{Gw170817: Observation of gravitational waves from a binary
  neutron star inspiral},
  \href{https://doi.org/10.1103/PhysRevLett.119.161101}{\emph{Phys. Rev. Lett.}
  {\bfseries 119} (Oct, 2017) 161101}.

\bibitem{Abbott_2017}
B.~P. Abbott, R.~Abbott, T.~D. Abbott, F.~Acernese, K.~Ackley, C.~Adams et~al.,
  \emph{Gravitational waves and gamma-rays from a binary neutron star merger:
  {GW}170817 and {GRB} 170817a},
  \href{https://doi.org/10.3847/2041-8213/aa920c}{\emph{The Astrophysical
  Journal} {\bfseries 848} (oct, 2017) L13}.

\bibitem{schutz1977variational}
B.~F. Schutz and R.~Sorkin, \emph{Variational aspects of relativistic field
  theories, with application to perfect fluids}, {\emph{Annals of Physics}
  {\bfseries 107} (1977) 1--43}.

\bibitem{Felice_2016}
A.~D. Felice and S.~Mukohyama, \emph{Phenomenology in minimal theory of massive
  gravity}, \href{https://doi.org/10.1088/1475-7516/2016/04/028}{\emph{Journal
  of Cosmology and Astroparticle Physics} {\bfseries 2016} (apr, 2016)
  028--028}.

\end{thebibliography}\endgroup
\bibliographystyle{JHEP}

\end{document}